\documentclass[
aip,
author-year,
onecolumn,
notitlepage,
superscriptaddress,
12pt]{revtex4-2}

\usepackage[margin=1.2in]{geometry}                \usepackage{microtype}                     \usepackage{mathpazo}                      

\usepackage{amsfonts, amsmath, bm, bbm, amssymb, amsthm, bbold, mathrsfs}
\usepackage{graphicx}\usepackage{dcolumn}\usepackage[dvipsnames]{xcolor}
\usepackage{url}
\usepackage{hyperref}
\hypersetup{hidelinks}

\usepackage{algorithm}
\usepackage{caption}
\usepackage{algpseudocode}

\newcommand{\abs}[1]{\left\lvert{#1}\right\rvert}

\newcommand{\myMSC}[1]{\par\noindent
  \textit{2020 Mathematics Subject Classification:} #1
}

\usepackage{cancel}
\usepackage[normalem]{ulem}
\usepackage[weather]{ifsym}
\usepackage{physics}

\newcommand{\mean}[2]{\mathbb{E}_{#1}[#2]}
\newcommand{\cmean}[3]{\mathbb{E}_{#1}[#3\vert #2]}

\newcommand{\Var}[1]{\mathrm{\mathbb{V}ar}[#1]}

\newcommand{\Vart}[2]{\mathrm{\mathbb{V}ar}_{#2}[#1]}
\newcommand{\Covt}[2]{\mathrm{\mathbb{C}ov}_{#2}[#1]}
\newcommand{\indicator}[2]{\mathbb{1}_{#1}[#2]}

\newcommand{\springeronly}[1]{}
\newcommand{\arxivonly}[1]{#1}
\newcommand{\springerline}{}

\newcommand{\paperabstract}[1]{\begin{abstract}#1\end{abstract}}

\begin{document}

\title{Slice Monte Carlo Integration}

\author{Johannes K. Krondorfer}
\email{johannes.krondorfer@gmail.com}
\affiliation{Institute of Experimental Physics, Graz University of Technology, Petersgasse 16, 8010 Graz, Austria}

\author{Christian W. Binder}
\affiliation{Department of Physics, University of Oxford, Parks Rd, Oxford OX1 3PJ, United Kingdom}

\author{Matthias Neumann}
\affiliation{Institute of Statistics, Graz University of Technology, Kopernikusgasse 24/III, 8010 Graz, Austria}
\affiliation{GraML -- Graz Center for Machine Learning, Rechbauerstraße 12, 8010, Graz, Austria}

\author{Wolfgang von der Linden}
\affiliation{GraML -- Graz Center for Machine Learning, Rechbauerstraße 12, 8010, Graz, Austria}
\affiliation{Institute of Theoretical and Computational Physics, Graz University of Technology, Petersgasse 16, 8010 Graz, Austria}

\date{\today}

\paperabstract{
Numerical integration involving expensive target functions is a common bottleneck in Bayesian inference and simulation. When a cheap surrogate is available, standard approaches such as reweighting or importance sampling often suffer from high variance and inefficient use of function evaluations. We introduce Slice Monte Carlo integration (S$\ell$MC), a method that leverages a Nested Sampling-like procedure on the surrogate to partition the space into informative strata, or \textit{slices}, while generating samples in the parameter space drawn from the prior within each slice. This enables stratified Monte Carlo integration of the expensive target function over the surrogate-induced partition, yielding an efficient estimate of the target integral. The surrogate level sets therefore make the induced ordering of the parameter space the relevant information for S$\ell$MC, rather than the pointwise target-to-surrogate ratios governing importance-sampling efficiency. Another key advantage of S$\ell$MC is the decoupling of slice volume estimation from the evaluation of the target, allowing the refinement of the slice-volume estimates without additional target evaluations. We investigate the properties of S$\ell$MC, demonstrate how to efficiently generate posterior samples, and assess its robustness in controlled Gaussian benchmark families with continuously tunable surrogate-target mismatch.
}


\maketitle
\myMSC{65C05; 65D30}

\section{Introduction}
Efficient numerical integration is a key challenge in many disciplines, including data science, physics, engineering, and Bayesian statistics, where high-dimensional integrals arise in tasks such as model selection~\citep{gelman2014understanding, goggans2013}, parameter estimation and uncertainty quantification~\citep{andrieu2003introduction, jasa2012, veitch2015, abbott2019, najm2009, Yun2016uq}, and partition function evaluation~\citep{partay2014nested, partay2021, szekeres2018direct, binder2022ridge, bolhuis2018}. While classical quadrature methods quickly become intractable in high dimensions, stochastic methods such as Monte Carlo integration, importance sampling~\citep{lepage1978, press1990, caflisch1998,lepage2021}, and Nested Sampling~\citep{skilling2004, skilling2006} have proven effective when the integrand is inexpensive to evaluate.

In many scientific and engineering applications, however, evaluating the target function is computationally expensive. A single evaluation may require a large-scale simulation, the solution of a complex physical model, or the processing of high-resolution data~\citep{najm2009, partay2021, veitch2015}. In such settings, standard integration methods can become inefficient because they require many target evaluations. At the same time, low-fidelity approximations or surrogate models are often available and capture useful structure of the integrand at substantially lower cost. These approximations may take the form of coarse physical models~\citep{najm2009, Yun2016uq}, simplified energy landscapes~\citep{partay2021, binder2022ridge}, or machine-learned regressors trained on high-fidelity data~\citep{Lu2019surr, krondorfer2023gpr, kovacs2025maceoff, uteva2018, Lu2022}. The central challenge is to use such approximations to guide integration, accelerate computation, and reduce variance without introducing bias.

Several approaches have been proposed to address this challenge. Sequential Monte Carlo methods construct a sequence of weighted particle approximations to the target distribution, often combining reweighting, resampling, and Markov chain Monte Carlo updates to balance bias and variance~\citep{salomone2025smc, wills2023smc}. Multifidelity Monte Carlo and Multilevel Monte Carlo methods combine simulations at different fidelities using control variates or telescoping estimators to reduce variance while maintaining unbiasedness~\citep{peherstorfer2016mfmc, peherstorfer2018mfmc, Giles2015mlmc}. Adaptive importance sampling techniques construct proposal distributions informed by surrogate models or auxiliary densities to concentrate samples in high-contribution regions~\citep{lepage1978, tokdar2010is}. Stratified sampling methods reduce variance by partitioning the parameter space into regions of approximately equal weight. However, standard stratification often relies on simple strata, such as axis-aligned partitions~\citep{press1990}, which may poorly reflect complex posterior geometries. While these methods can perform well for expectation estimation, they often struggle to scale efficiently to high-dimensional or multimodal distributions, and typically do not yield posterior samples without additional approximation or resampling techniques~\citep{ashton2022,feroz2009}.

In contrast, Nested Sampling provides a framework for estimating Bayesian evidence and generating posterior samples simultaneously~\citep{skilling2004, skilling2006}. By constructing nested likelihood contours and estimating the associated constrained prior masses, it enables efficient exploration of complicated multimodal distributions. Recent extensions incorporate dynamic allocation of samples~\citep{Higson2019DynamicNS, dynesty2020}, slice-sampling-based exploration of the constrained prior in high dimensions~\citep{handley2015polychord}, or adaptations for multimodality~\citep{feroz2009, Buchner2023ns, buchner2021ultranest}, but they still rely on target evaluations to estimate constrained prior masses. Moreover, reweighting schemes based on Nested Sampling still require many target evaluations, because the fidelity of the volume estimate remains directly tied to the number of target function calls, as we discuss in Section~\ref{sec:RW}.

In this paper, we introduce Slice Monte Carlo integration (S$\ell$MC), a method that combines the strengths of stratified Monte Carlo and Nested Sampling for surrogate-based integration. S$\ell$MC applies a Nested Sampling-like procedure to the surrogate to partition the space into informative strata, or \textit{slices}, while generating prior samples within each slice. This enables stratified Monte Carlo integration over the surrogate-induced partition and yields efficient estimates of the target integral. The surrogate level sets therefore make the induced ordering of parameter space the relevant information for S$\ell$MC, rather than the pointwise target-to-surrogate ratios governing importance-sampling efficiency. A further key advantage over standard reweighting techniques is that volume estimation is decoupled from target function evaluation. Slice volumes can therefore be estimated accurately using the cheap surrogate, while the expensive target-evaluation budget can be chosen independently. Furthermore, the surrogate-induced partition facilitates efficient generation of posterior samples via rejection sampling.

The term slice in this paper denotes surrogate-induced strata for integration. This differs from Neal's slice sampling for Markov chain Monte Carlo~\citep{neal2003slice}, and from the use of slice sampling as a constrained-prior sampler in NS implementations such as PolyChord~\citep{handley2015polychord}.
The resulting level-set construction is closely related to the sequential Monte Carlo formulation of Nested Sampling by \citet{salomone2025smc}. In the terminology used here, the likelihood-driven setting corresponds to the case in which target and surrogate are identical, and the analysis treats the resulting scheme when Markov chain Monte Carlo is used to generate constrained samples. The surrogate-informed setting raises a distinct and practically important question: how efficiently and robustly can an imperfect surrogate guide integration of an expensive target function?

The remainder of this paper is structured as follows. In Section~\ref{sec:background}, we formalize the problem setup and review stochastic integration methods relevant to the development of S$\ell$MC. Section~\ref{sec:slim} introduces the slicing framework, followed by a detailed presentation of the S$\ell$MC algorithm, including its theoretical properties, such as unbiasedness and variance, and a discussion on generating posterior samples. Finally, Section~\ref{sec:experiments} demonstrates the performance of the proposed method on benchmark problems.

\section{Problem Setup and Related Methods}\label{sec:background}
We are interested in evaluating integrals of the form
\begin{equation}\label{eq:evidence}
    \mathcal{Z} = \mathbb{E}_\pi[L] = \int_{\Omega} L(\theta)\;\mathrm{d}\pi(\theta)\,,
\end{equation}
over a Borel-measurable parameter space $\Omega \subset \mathbb{R}^d$ for $d \geq 1$, where $L: \Omega \rightarrow \mathbb{R}$ is the target function and $\pi$ is a prior probability distribution on $\Omega$. Specifically, this includes evidence integrals in the context of Bayesian inference, where $L$ is the non-negative likelihood function or partition function in statistical physics, e.g., the non-negative exponential Boltzmann factor. This integral is particularly demanding because the prior probability distribution function $\pi(\theta)$ typically has a much smaller variability than the target function $L(\theta)$.

Another goal of this paper is based on the posterior distribution
$\mathrm{d}p(\theta) = L(\theta)\,\mathrm{d}\pi(\theta) / \mathcal{Z}$, for which either the generation of uncorrelated samples or posterior expectation values of the form
\begin{equation}\label{eq:expectation}
    \mathcal{G} = \mathbb{E}_{p}[g] = \int_{\Omega} \; g(\theta)\;\mathrm{d} p(\theta)\,.
\end{equation} 
are of great interest. This type of problem arises in many contexts, including statistical inference, uncertainty quantification, and statistical physics, where, e.g., the posterior expectation corresponds to the thermodynamic expectation of some physical observable.

To estimate such integrals and obtain samples, various stochastic integration techniques have been developed. Below, we briefly review the most relevant integration approaches that form the conceptual foundation for Slice Monte Carlo integration (S$\ell$MC) proposed in the present paper.

\subsection{Monte Carlo Integration}\label{sec:MC}
In standard Monte Carlo integration (simple sampling), expectations with respect to $\pi$  are estimated using the empirical mean over independent and identically distributed samples $\Theta_1, \dots, \Theta_N \overset{\mathrm{iid}}{\sim} \pi$. For the evidence integral $\mathcal{Z} = \mathbb{E}_{\pi}[L]$, this yields the estimator
\begin{equation}\label{eq:MC Z}
    \widehat{\mathcal{Z}}_N^\text{MC} = \frac{1}{N}\sum_{i=1}^N L(\Theta_i)\,,
\end{equation}
which is unbiased, converges at rate $\mathcal{O}(N^{-1/2})$ and is asymptotically normal provided that $\mathrm{\mathbb{V}ar}_{\pi}[L] < \infty$~\citep{caflisch1998}. Its variance is given by
\begin{equation}\label{eq:MC var Z}
    \mathrm{\mathbb{V}ar}[\widehat{\mathcal{Z}}_N^\text{MC}] = \frac{1}{N} \mathrm{\mathbb{V}ar}_{\pi}[L]\,.
\end{equation}

\subsubsection{Monte Carlo Importance Sampling}\label{sec:IS}
Simple sampling may be inefficient when significant contributions to $\mathcal{Z}$ come from regions of low probability under $\pi$. If a nonnegative surrogate function $\phi(\theta)$ can be found for which the integral 
$\mathcal{Z}_\phi = \mathbb{E}_{\pi}[\phi] $ is known, or can easily be evaluated to arbitrary precision, then one can introduce a new probability measure $\mathrm{d}p_\phi(\theta) = \phi(\theta)\,\mathrm{d}\pi(\theta) / \mathcal{Z}_\phi$ and the target integral $\mathcal{Z}$ can be written as
\begin{equation}\label{eq:IS int_new}
    \mathcal{Z} = \mathcal{Z}_\phi \;\mathbb{E}_{p_\phi}[L / \phi]\,.
\end{equation}

The resulting Monte Carlo estimator, which we refer to as importance sampling (IS) estimator, remains unbiased and has variance
\begin{equation}\label{eq:IS var Z}
    \mathrm{\mathbb{V}ar}[\widehat{\mathcal{Z}}_N^\text{IS}] = \frac{\mathcal{Z}_\phi^2}{N} \mathrm{\mathbb{V}ar}_{p_\phi}[L / \phi]\,.
\end{equation}
Crucially, $\mathcal{Z}_\phi$ must be known, which in itself can be a complicated task. If the surrogate $\phi(\theta)$ resembles closely the target function $L(\theta)$ then the numerical effort to compute $\mathcal{Z}_\phi$ is the same as the numerical effort to compute $\mathcal{Z}$, unless it allows an analytical evaluation, or to use more efficient numerical techniques which are not applicable to $L(\theta)$. 
However, since the estimator depends on the pointwise ratio $L/\phi$, mismatch between $\phi$ and $L$ can lead to large variance.
Adaptive schemes that iteratively construct surrogates, e.g. product approximations $\phi(\theta) = \prod_{j=1}^d g_j(\theta_j)$, have been proposed~\citep{lepage1978, lepage2021}, but typically rely on independence assumptions across dimensions.

\subsubsection{Stratified Monte Carlo Integration}\label{sec:strat MC}
Stratification can also reduce the variance of the integrand. To that purpose, $\Omega$ is partitioned into disjoint regions (strata) $\mathscr{S}_1, \dots, \mathscr{S}_{M}$, with $\Omega = \cup_{i=1}^M \mathscr{S}_{i}$,
and the integral can be written as
\begin{equation}
    \mathcal{Z} = \sum_{i=1}^{M} \int_{\mathscr{S}_i} L(\theta)\;\mathrm{d}\pi(\theta)\,.
\end{equation}
Each term is then estimated separately via Monte Carlo
\begin{equation}
    \widehat{\mathcal{Z}}_\mathscr{S} = \sum_{i=1}^{M} \frac{S_i}{N_i} \sum_{k=1}^{N_i} L(\Theta_{i,k})\,,
\end{equation}
where $S_i = \pi[\mathscr{S}_i]$ is the prior mass of stratum $\mathscr{S}_i$, and $\Theta_{i,k} \overset{iid}{\sim}\pi|_{\mathscr{S}_i}$ for $i = 1,\ldots, M$ and $k = 1, \ldots, N_i$.

Sample allocation can be optimized to minimize the total variance. Writing $N = \sum_{i=1}^M N_i$, the optimal allocation is~\citep{press1990}
\begin{equation}\label{eq:MC strat number of sample points}
    \frac{N_i}{N} = \frac{S_i \sigma_{\mathscr{S}_i}}{ \sum_{k=1}^M S_k \sigma_{\mathscr{S}_k}}\,,
\end{equation}
where $\sigma^2_{\mathscr{S}_i} = \mathrm{\mathbb{V}ar}_\pi[L(\Theta) \vert \Theta\!\in\!\mathscr{S}_i]$ denotes the variance of $L$ in stratum~$\mathscr{S}_i$.
So the minimal variance, if only the number of points per stratum $N_1, \ldots, N_M$ are optimized, reads
\begin{equation}
\Vart{\widehat{\mathcal{Z}}_\mathscr{S}}{\text{opt}}
     = \frac{1}{N} \qty(\sum_{i=1}^M   S_i \sigma_{\mathscr{S}_i})^{\!\!2}
\end{equation}
as compared to
\begin{equation}
\Vart{\widehat{\mathcal{Z}}_\mathscr{S}}{\text{eq}}
     = \frac{M}{N} \sum_{i=1}^M   S^2_i \sigma_{\mathscr{S}_i}^2
\end{equation}
if the same number $N_i = N/M$ is used in all strata.
The difference
\begin{align}
\springeronly{\begin{split}&}
\Vart{\widehat{\mathcal{Z}}_\mathscr{S}}{\text{eq}} - \Vart{\widehat{\mathcal{Z}}_\mathscr{S}}{\text{opt}} \springeronly{\\} &= \frac{M}{N}
\bigg( \sum_{i=1}^M   
\big[S_i \sigma_{\mathscr{S}_i} - \overline{S\sigma }\big]^2\bigg)
> 0\;,\springerline\quad \text{with } \overline{S\sigma} :=
\frac{1}{M} \sum_{i=1}^M   S_i \sigma_{\mathscr{S}_i}
\springeronly{\end{split}}
\end{align}

Despite its potential, stratification poses two key challenges: identifying effective strata \textit{a priori} is nontrivial in high dimensions, and sampling from the prior within each stratum may be difficult and computationally expensive. Common approaches use axis-aligned partitions or recursive bisection along directions of high variance~\citep{press1990,lepage2021}. Adaptive stratification can improve performance, but a large number of evaluations may still be needed. Additionally, simple bisections of dimensions can violate symmetries of the integrand and therefore reduce the efficiency. This limitation is addressed in S$\ell$MC, where we design informative strata that are adapted to the target function, via a Nested Sampling-like procedure applied to a surrogate.

\subsection{Nested Sampling}\label{sec:NS}
\subsubsection{Rewriting the Evidence Integral}
Before introducing S$\ell$MC, we briefly review the principles of Nested Sampling (NS). NS begins by reformulating the evidence integral in Equation~\eqref{eq:evidence} over a positive likelihood function $L:\Omega\to\mathbb{R}_{>0}$ as an integral over the constrained prior mass $X(\lambda) := \pi[L(\Theta) > \lambda]$
\begin{equation}
    \mathcal{Z} = \int_0^1 \widetilde{L}(x)\;\mathrm{d}x\,,
\end{equation}
where
\begin{equation}
    \widetilde{L}(x) = \sup\{\lambda\in\mathbb{R} \,\vert\, X(\lambda) > x\}
\end{equation}
is the generalized inverse of $X$, also interpreted as the likelihood corresponding to the contour enclosing prior mass~$x$~\citep{skilling2004, skilling2006,
vdLindenDosevToussaint2014, Fowlie_2021, schittenhelm2021nestedsamplinglikelihoodplateaus}.

This transforms the high-dimensional integral into a one-dimensional integral over the unit interval, which can then be approximated using standard quadrature
\begin{equation}\label{eq:NS estimator}
    \mathcal{Z} \approx Z = \sum_{i=1}^K w_i \widetilde{L}_i\,,
\end{equation}
where $w_i = X_{i-1} - X_i$ are weights associated with prior mass differences and $\widetilde{L}_i$ are corresponding likelihood values. 

Since $\widetilde{L}(x)$ is a monotonically decreasing function in the constrained prior mass $x$, Equation~\eqref{eq:NS estimator} yields an upper bound for the true $\mathcal{Z}$, while $\sum_{i=1}^K w_i \widetilde{L}_{i-1}$ is a strict lower bound. However, evaluating $X(\lambda)$ exactly is intractable, and it must be approximated stochastically~\citep{skilling2004, skilling2006}.

\subsubsection{The Nested Sampling Procedure}
NS estimates $X$ by random sampling from level sets of increasing likelihood. The key identity is that for $\Theta \sim \pi|_{>L^*}$, i.e. drawn from the prior constrained to the set $\{L > L^*\}$, $X(L(\Theta)) \sim \mathrm{Uniform}(0, X^*)$, where $X^* = X(L^*)$. This identity holds if $X$ is continuous, which is the case if $L$ has no plateaus of non-zero probability with respect to~$\pi$~\citep{Fowlie_2021, schittenhelm2021nestedsamplinglikelihoodplateaus}.

By drawing $N_{\mathrm{live}}$ independent samples from $\pi\vert_{>L^*}$ and sorting them by likelihood,
\begin{equation}\label{eq:ordered sample}
\begin{array}{ccccccc}
     L_{(1)} &>& L_{(2)} &>& \cdots &>& L_{(N_{\mathrm{live}})}  \\
     X_{(1)} &<& X_{(2)} &<& \cdots &<& X_{(N_{\mathrm{live}})}\,,
\end{array}
\end{equation}
the smallest likelihood corresponds to a prior mass $X_{(N_{\mathrm{live}})} = t X^*$ where $t \sim \mathrm{Beta}(N_{\mathrm{live}},1)$, as given by the maximum statistics~\citep{skilling2006,vdLindenDosevToussaint2014}.
The number $N_{\mathrm{live}}$ is typically referred to as the number of live points.

This procedure can be used to compress the prior mass iteratively, via the recursion
\begin{equation}
    \springeronly{\begin{aligned}&}
    X_i = t_i X_{i-1}, \springerline\quad \text{with } t_i \sim \mathrm{Beta}(N_{\mathrm{live}},1), \quad X_0 = 1\,,
    \springeronly{\end{aligned}}
\end{equation}
The contraction factors $t_i$ are mutually independent, as they are associated with independent uniform samples from the constrained prior. Conditional on the sampled likelihood values $\{\widetilde L_i\}$, the quadrature estimate is therefore a function of the compression factors, and we write $Z=Z(\bm t)$ to indicate this dependence, where $\bm{t} = (t_1, \dots, t_K)$ denotes the vector of compression factors.

At each step, the live point with the lowest likelihood is discarded and stored, and a new point is drawn from the new constrained prior $\pi\vert_{>L_i}$. In practice the constrained draw is the dominant cost of NS, and a variety of constrained samplers have been proposed, ranging from standard Markov chain Monte Carlo methods~\citep{skilling2006} and ellipsoidal rejection~\citep{feroz2009} to slice-sampling chains~\citep{handley2015polychord}.
The other live points already fulfill the constraint by construction. Repeating this for $K$ steps yields a sequence $\{\theta_i, \widetilde{L}_i, X_i(\bm{t})\}$ which forms the quadrature rule in Equation~\eqref{eq:NS estimator}. The use of the maximum statistic in each iteration is optimal in nested sampling, as it minimizes the estimator variance and maximizes the resolution of the quadrature~\citep{skilling2006}. While alternative order statistics are theoretically possible, they generally lead to inferior performance. To ensure that the quadrature adequately captures the bulk of the posterior mass, the number of iterations should satisfy $K\geq N_{\mathrm{live}}H$, where $H$ is the relative entropy (or Kullback–Leibler divergence) from the posterior to the prior~\citep{skilling2006}. This condition ensures that the nested sampling trajectory penetrates deeply enough into high-likelihood regions to estimate both the evidence and the posterior accurately.

Dynamic nested sampling methods~\citep{Higson2019DynamicNS, dynesty2020} extend this framework by adaptively varying the number of points and termination criteria during sampling to enhance efficiency, particularly in the estimation of posterior expectations.

\subsubsection{Error analysis}
NS in this form produces not only a point estimate but a distribution over evidence estimates $Z(\bm t)$, conditional on the sampled likelihood values, which we indicate by conditioning on $\bm\theta$. The remaining uncertainty in $Z(\bm t)$ then comes from the stochastic compression factors $\bm t$. The compression factors may be set deterministically, e.g., $t_i = \exp(-1/N_{\mathrm{live}})$ or $t_i = N_{\mathrm{live}}/(N_{\mathrm{live}}+1)$, or sampled from $\mathrm{Beta}(N_{\mathrm{live}},1)$~\citep{skilling2006, keeton2011, vdLindenDosevToussaint2014,walter2017point, Higson2018error}.

From now on, we assume that $t_1, \ldots, t_K$ are iid with $t_1 \sim \mathrm{Beta}(N_{\mathrm{live}},1)$. Error analysis can be conducted either using information-theoretic arguments~\citep{skilling2006}, or through a moment-based analysis~\citep{keeton2011,vdLindenDosevToussaint2014}. The latter approach yields formulas for the expectation and variance of $Z(\bm{t})$, given the knowledge of all sampled points during the procedure. Exploiting the independence of the compression factors $t_i$ and using $\mathbb{E}[t_i] = N_{\mathrm{live}}/(N_{\mathrm{live}}+1)$, one obtains
\begin{equation}
    \mathbb{E}[Z\vert\bm{\theta}]\! =\! \sum_{i=1}^K \mathbb{E}[w_i\vert\bm{\theta}]\, \widetilde{L}_i = \frac{1}{N_{\mathrm{live}}}\!\sum_{i=1}^K \! \!\qty(\frac{N_{\mathrm{live}}}{N_{\mathrm{live}}+1})^{\!\!i} \widetilde{L}_i\,.
\end{equation}
In practical applications, it is often sufficient to consider the distribution $Z(\bm{t})$ given the observed data, whose uncertainty arises from the stochastic nature of NS. Taking its conditional expectation yields a Bayesian point estimator of the evidence. It can be shown that this point estimator is asymptotically unbiased and asymptotically normally distributed under additional regularity conditions~\citep{chopin2010}. Note, however, that the remaining randomness of the point estimator is contained in $\widetilde{L}_i$, which are obtained by the sorting procedure, and therefore, a simple error analysis of the point estimator is not applicable. Nevertheless, reformulations in terms of sequential Monte Carlo algorithms exist, which provide an inherently unbiased estimator for Nested Sampling~\citep{salomone2025smc}. For high-dimensional integrals, Skilling pointed out that instead of $Z$ the relevant quantity is $\log(Z)$, which is usually closer to a normal distribution~\citep{skilling2009convergence, vdLindenDosevToussaint2014}.

For the conditional variance, \citet{keeton2011} and \citet{vdLindenDosevToussaint2014} provide an explicit expression when the number of live points is constant. Here we will only investigate the general structure of the variance, to analyze the sources of uncertainty,
\begin{align}
\begin{split}
    \mathrm{\mathbb{V}ar}[Z\vert\bm{\theta}] &= \mathrm{\mathbb{V}ar}[\sum_{i=1}^K w_i \widetilde{L}_i\vert\bm{\theta}]\springerline= \sum_{i=1}^K \widetilde{L}_i^2 \mathrm{\mathbb{V}ar}[w_i\vert\bm{\theta}] \springerline + 2\!\sum_{\substack{i,j =1 \\ i<j}}^K\! \widetilde{L}_i \widetilde{L}_j \mathrm{\mathbb{C}ov}[w_i,w_j\vert\bm{\theta}]\,.
\end{split}
\end{align}
Using the decomposition $w_j = X_i \prod_{k=i+1}^{j-1} t_k\; (1-t_j)$ and the independence of the compression factors, we get
\begin{align}\label{eq:NS var}
\begin{split}\springeronly{&}
    \mathrm{\mathbb{V}ar}[Z\vert\bm{\theta}] \springerline =\sum_{i=1}^K\!\Bigg(\widetilde{L}_i^2\arxivonly{&} \mathrm{\mathbb{V}ar}[w_i\vert \bm{\theta}] \springerline + 2\widetilde{L}_i \mathrm{\mathbb{C}ov}[w_i,X_i\vert\bm{\theta}]\springerline\!\sum_{j=i+1}^K\!\! \Big(\widetilde{L}_j \mathbb{E}[1\!-\!t_j]\prod_{k=i+1}^{j-1}\!\!\!\mathbb{E}[t_k] \Big) \Bigg)\,.
\end{split}
\end{align}
We will see in Section~\ref{sec:SlMC}, why this particular decomposition of the variance of NS allows for an easy interpretation of the variance terms present in S$\ell$MC. Further details can also be found in~\citep{vdLindenDosevToussaint2014}, where it is also shown that for a large number of live points $N_{\mathrm{live}}$, the variance of $Z$ is $\order{1/N_{\mathrm{live}}}$ like in standard MC. This implies that increasing the number of live points reduces the uncertainty, but at a price: The mean compression factor for large $N_{\mathrm{live}}$ is  $t \approx e^{-1/N_{\mathrm{live}}}$, hence the mean prior mass at step $i$ is $X_i \approx e^{-i/N_{\mathrm{live}}}$.
In realistic problems, the dominant part of the likelihood corresponds to a very small prior mass $X^*$, which has to be reached for convergence. The number of necessary steps $m^*$ follows from
$X_{m^*} = X^*$ and gives $m^* = N_{\mathrm{live}} \abs{\ln(X^*)}$. Hence, the number of steps and therefore the number of function evaluations increases proportionally to $N_{\mathrm{live}}$ and in turn inversely proportional to $\varepsilon_Z^2$, the desired relative uncertainty.
We see, the stochastic uncertainty in NS comes entirely from the constrained prior masses, which are determined by $N_{\mathrm{live}}$. We will see later that it is possible to determine the thresholds differently and obtain a way to strongly reduce the uncertainty of the constrained prior masses by adding uncertainty to the likelihood values, which allows for more flexibility to optimize the algorithm.

\subsubsection{Posterior Representatives}
NS can also be used to generate posterior representatives. Since the weighted likelihood values $\widetilde{L}_i$ are associated with constrained prior masses $X_i$, one can assign normalized weights
\begin{equation}
    p_i(\bm{t}) = w_i(\bm{t}) \widetilde{L}_i / Z(\bm{t})
\end{equation}
to obtain a discrete approximation of the posterior distribution $\mathrm{d}p(\theta) \!\!=\! L(\theta)\,\mathrm{d}\pi(\theta) / \!\mathcal{Z} $. This still contains the uncertainty of the compression factors $\bm{t}$. The probability $\hat p_i$ for $\theta_i$ is now given by the mean $\mean{\bm{t}}{w_i(\bm{t}) \widetilde{L}_i / Z(\bm{t})}$, which is easily obtained by MC means.
This can be used to estimate the expectation $\mathcal{G} := \mathbb{E}_{p}[g]$ of an $p$-integrable quantity $g:\Omega\to\mathbb{R}$ by 
\begin{equation}
    G(\bm{t}) = \sum_{i=1}^K p_i(\bm{t}) g(\theta_i)
\end{equation}
or to obtain an equally weighted posterior sample by accepting each $\theta_i$ with probability $\hat p_i / C$, with $C \geq \max_i \hat p_i$~\citep{skilling2006}.

Similarly, a sampling-based or moment-based error analysis is also possible in this case. However, it is unclear whether $g(\theta_i)$ is indeed a good representative of the function $g$ within the volume associated with the weight $w_i$.\citep{Higson2018error}

\subsection{Reweighting}\label{sec:RW}
Similar to importance sampling, the NS procedure can be reused to estimate the integral of a target function $L$ using a surrogate model $\phi$~\citep{cameron2014recursive, chopin2010, feroz2019importanceINS}. Specifically, if samples $\theta_i^{\phi}$ are drawn using NS applied to $\phi$, we can express the evidence $Z$ for $L$ as the expectation in Equation~\eqref{eq:IS int_new}, yielding
\begin{equation}\label{eq:Z}
    \springeronly{\begin{aligned}&}
    Z(\bm{t}_\phi) \springerline = \sum_{i=1}^K p_{\phi,i}(\bm{t}_\phi)\, Z_\phi(\bm{t}_\phi) \, L(\theta_{\phi,i}) / \widetilde{\phi}_i \springerline = \sum_{i=1}^K w_{\phi,i}(\bm{t}_\phi) \, L(\theta_{\phi,i})\,,
    \springeronly{\end{aligned}}
\end{equation}
where $p_{\phi,i} =  w_{\phi,i} \widetilde{\phi}_i / Z_\phi$ are posterior representatives of the surrogate model, and $w_{\phi,i}$ the volume elements derived from the surrogate model. 
As a matter of fact, $w_{\phi,i}(\bm{t}_\phi)$ is independent of the likelihood and it therefore has the same PDF as $w_i(\bm{t})$.
The estimator in Equation~\eqref{eq:Z} is analogous to that of standard NS for $Z$, but here $L$ is evaluated at the sampled points $\theta_{\phi,i}$ obtained by NS based on $\phi$.

The surrogate-based quadrature inherits the volume uncertainty of NS applied to $\phi$. In addition, if $\phi$ and $L$ do not induce the same ordering, the value $L(\theta_{\phi,i})$ may be a poor representative of the target function over the prior-mass element $w_{\phi,i}$, which introduces an additional source of variance.

At first glance, Equation~\eqref{eq:Z} looks like there is nothing gained by using the surrogate. One important advantage, however, is the fact that the generation of constrained prior points, which typically necessitates several likelihood evaluations~\citep{skilling2006,handley2015polychord}, is done with the low-cost surrogate. And that is, as we will see in the results section, a huge advantage if the true likelihood is expensive.

A key advantage of reweighting over classical importance sampling is that it becomes NS when $L$ and $\phi$ induce the same ordering over $\Omega$, i.e. if $L(\theta_1) < L(\theta_2)$ is equivalent to $\phi(\theta_1) < \phi(\theta_2)$
for all $\theta_1, \theta_2 \in \Omega.$
In this case, the nested samples $\theta_{\phi,i}$ and associated volumes $w_{\phi,i}$ exactly match those that would have been produced by applying NS directly to $L$, and the reweighted estimator reproduces the NS result.

As in standard NS, the stochasticity lies in the volume elements $w_{\phi,i}$, so moment-based error analysis~\citep{keeton2011} still applies. However, a new source of error arises, as $L$ may vary significantly within the region associated with $w_{\phi,i}$, making $L(\theta_{\phi,i})$ a potentially poor representative of that volume. Additionally, the number of point evaluations of $L$ must match the number of estimated volumes $w_{\phi,i}$, tying computational cost to quadrature resolution. Both issues are directly addressed by S$\ell$MC, which decouples volume estimation from point evaluations for the target function and allows the within-slice target-evaluation budget to be chosen independently.

\section{Slicing and Slice Monte Carlo}\label{sec:slim}

\subsection{Slicing \--- Partitioning of the Parameter Space}\label{sec:slicing}
With these prerequisites in place, we now introduce the partitioning of the parameter space into slices and the corresponding slice Monte Carlo integration (S$\ell$MC), which combines ideas of Nested Sampling, reweighting, and stratified Monte Carlo. By using a positive function $\phi:\Omega\to\mathbb{R}_{>0}$, which can be chosen as a cheap surrogate for the target function $L$, the core idea is to construct a sequence of thresholds $\{\phi_i\}_{i=0}^{M-1}$, with $\phi_0 = 0$, using a Nested Sampling-like procedure. These thresholds define level sets 
\begin{equation}
    \springeronly{\begin{aligned}&}
    \mathscr{X}_i := \{\theta\in\Omega\,\vert\, \phi(\theta) > \phi_i\}\,,\springerline\quad X_i := \pi[\mathscr{X}_i]\,,\quad i=0,\dots,M-1\,.
    \springeronly{\end{aligned}}
\end{equation}
where $X_i$ is the constrained prior mass above $\phi_i$. For notational convenience, we set $\mathscr{X}_M=\emptyset$ and $X_M=0$. This induces a partition of the parameter space into slices 
\begin{equation}
\springeronly{\begin{aligned}&}
(\mathscr{S}_i, S_i) = (\mathscr{X}_{i-1} \setminus \mathscr{X}_i,\; X_{i-1} - X_i)\,,\springerline\quad i=1,\dots,M\,,
\springeronly{\end{aligned}}
\end{equation}
where $S_i = \pi[\mathscr{S}_i]$ is the prior mass in slice $\mathscr{S}_i$.
After constructing the slices, we can use samples drawn from the constrained prior within each slice $\mathscr{S}_i$ to compute the target integral \textit{via} stratified Monte Carlo integration.

Constrained sampling within each slice is generally very demanding, due to two constraints -- upper and lower --  on the value of the slicing function $\phi$.
However, the Nested Sampling-like procedure, where only the lower constraint is used to construct the slices, can also be used to generate the necessary samples.

In each iteration $i=1,\dots,M-1$ a new threshold $\phi_i > \phi_{i-1}$ is chosen. It may be chosen arbitrarily, prescribed in advance, informed by previous runs, or determined by choosing the $k$-th order statistic of an iid sample $\Theta \sim \pi\vert_{\mathscr{X}_{i-1}}$ above the current threshold $\phi_{i-1}$. After the threshold is selected, $N_i$ independent iid samples are drawn from the constrained prior
\begin{equation}
    \Theta^{\mathscr{X}}_{i-1,1}, \dots, \Theta^{\mathscr{X}}_{i-1,N_i} \overset{\text{iid}}{\sim}\pi\vert_{\mathscr{X}_{i-1}}\,.
\end{equation}
Then the number $\hat n_i^\mathscr{X}$ of points satisfying $\phi(\Theta) > \phi_i$, i.e. 
\begin{equation}\label{eq:n X i}
    \hat n_i^\mathscr{X} :=\sum_{k=1}^{N_i} \mathbb{1}_{\mathscr{X}_i}(\Theta^{\mathscr{X}}_{i-1,k})\,,
\end{equation}
where $\mathbb{1}$ denotes the set indicator function,
is binomial, $\hat n_i^\mathscr{X}  \sim \mathrm{Binomial}(N_i,t_i)$. Here $t_i$ is the deterministic compression factor of prior mass $X_{i-1}$ to $X_{i}$, which coincides with the probability that a randomly selected point sampled from the constrained prior $\pi\vert_{\mathscr{X}_{i-1}}$ falls into the level set $\mathscr{X}_i$.
This allows for constructing a Monte Carlo estimator of the compression factor
\begin{equation}
    \hat t_i = \frac{\hat n^\mathscr{X}_i}{N_i} \,,
\end{equation}
leading to a sequence of estimators for the constrained prior masses
\begin{equation}
    \widehat{X}_i = \widehat{X}_{i-1} \hat t_i\quad\text{with} \quad \widehat{X}_0 = 1\,.
\end{equation}
The corresponding estimator for the prior mass contained in slice $i$ is estimated as $\widehat{S}_i = \widehat{X}_{i-1} - \widehat{X}_i$. Unrolling the recursion, this yields
\begin{align}
\begin{split}
    \widehat{X}_i = \prod_{k=1}^{i} \hat t_k\,,\quad\text{and}\quad
    \widehat{S}_i &= (1-\hat t_i)\prod_{k=1}^{i-1} \hat t_k\,,
\end{split}
\end{align}
with $\hat{t}_M  = t_M = 0$ and $\widehat{X}_M=X_M=0$ by convention.

For $i<M$, the binomial distribution of $\hat n_i^\mathscr{X}$ implies that the estimator of the compression factor is unbiased 
$\mathbb{E}[\hat t_i] = t_i$ and its variance is
\begin{equation}
    \Var{\hat t_i} =    \frac{\Var{\hat n_i^\mathscr{X}}}{N_i^2} = \frac{t_i  (1-t_i)}{N_i}\;.
\end{equation}
Thus, the relative uncertainty 
\begin{equation}
    \varepsilon_{t_i} :=\frac{\sqrt{\Var{\hat t_i}}}{t_i}  = \sqrt{\frac{1-t_i}{t_i N_i}}
\end{equation}
of the compression factors is proportional to $1/\sqrt{N_i}$. This, in turn, determines the relative uncertainty of the estimator of the constrained prior mass.
\begin{equation}\label{eq:unc:X}
    \varepsilon_{\widehat X_i}^2   = \sum_{k=1}^i   \varepsilon_{t_k}^2   + \order{\varepsilon_{t}^4}\;.
\end{equation}
A more detailed uncertainty analysis is provided in Section~\ref{sec:slmc var}.

Crucially, all points that fall within the slice $\mathscr{S}_i$, follow the prior distribution constrained to $\mathscr{S}_i$ , i.e. 
\begin{equation}
    \Theta \vert \Theta\!\in\!\mathscr{S}_i \sim \pi\vert_{\mathscr{S}_i}\quad\text{for}\quad \Theta\sim\pi\vert_{\mathscr{X}_{i-1}}\,.
\end{equation}
This allows to extract valid prior samples $\{\Theta^{\mathscr{S}}_{i,\ell}\}_{\ell=1}^{\hat n^\mathscr{S}_i}$ from each slice. Analogously to $\hat n_i^{\mathscr{X}}$ in Equation~\eqref{eq:n X i}, $\hat n_i^{\mathscr{S}}$ denotes the number of points falling into slice~$i$, with
\begin{equation}\label{eq:n S i}
\hat n_i^\mathscr{S} =
\begin{cases}
N_i-\hat n_i^\mathscr{X} = N_i(1-\hat t_i), & i < M,\\
\hat n_{M-1}^\mathscr{X} = N_{M-1}\hat t_{M-1}, & i = M.
\end{cases}
\end{equation}
Note that we will assume the thresholds and the number $N_{i}$ of points generated in $\mathscr{X}_{i-1}$ are chosen such that enough points fall into each slice, i.e. $\hat{n}_i^\mathscr{S}\gg 1$.

The complete procedure is summarized in Algorithm~\ref{alg:slimx}, which returns the constructed slices, the corresponding prior mass estimates, and a valid prior sample in each slice. 
\begin{algorithm}[H]
    \caption{Slicing}\label{alg:slimx}
    \begin{algorithmic}[1]
        \Require number of slices $M$
        \State Set $\widehat{X}_0 = X_0 = 1$, $\phi_0 = 0$ and $\mathscr{X}_0 = \Omega$
        \For{$i=1,\dots,M-1$}
            \State choose threshold $\phi_i > \phi_{i-1}$ as described in the text
            \State set $\mathscr{X}_i$ and $\mathscr{S}_i = \mathscr{X}_{i-1} \setminus \mathscr{X}_i$
            \State Sample $N_i$ points $\Theta^{\mathscr{X}}_{i-1,k}\sim\pi\vert_{\mathscr{X}_{i-1}}$ 
            \State determine sample $\mathscr{S}_i\supset\{\Theta^{\mathscr{S}}_{i,\ell}\} \subset \{\Theta^{\mathscr{X}}_{i-1,k}\}$
            \State compute compression $\hat t_i$ and $\widehat{X}_i = \hat t_i \widehat{X}_{i-1}$ and $\widehat{S}_i = \widehat{X}_{i-1} - \widehat{X}_i$
        \EndFor
        \State set $\mathscr{S}_M = \mathscr{X}_{M-1}$, $\widehat{S}_M = \widehat{X}_{M-1}$
        \State $\{\Theta^{\mathscr{X}}_{M-1,\ell}\} = \{\Theta^{\mathscr{S}}_{M,\ell}\} \springeronly{$\Statex\hspace{\algorithmicindent}$} = \{\Theta^{\mathscr{X}}_{M-2,k}\} \setminus \{ \Theta^{\mathscr{S}}_{M-1,\ell}\} $
        \Ensure $\{\mathscr{X}_i, \widehat{X}_i, \{\Theta^{\mathscr{X}}_{i,k}\}\}_{i=0}^{M-1}$, $\{\mathscr{S}_i, \widehat{S}_i, \{\Theta^{\mathscr{S}}_{i,\ell}\}\}_{i=1}^M$
    \end{algorithmic}
\end{algorithm}
This construction adopts a frequentist view of volume estimation. A Bayesian alternative would model the compression factors $t_i$ as Beta-distributed variables, based on the binomial sample. This would yield a hybrid strategy using Bayesian volume estimates while keeping Monte Carlo integration within each slice. In principle, a full Bayesian version is also possible if we can make prior assumptions on the target function within each slice.

\subsection{Slice Monte Carlo Integration}\label{sec:SlMC}
The slicing algorithm outlined above provides a partition $\mathscr{S}=(\mathscr{S}_1, \dots, \mathscr{S}_M)$ of the parameter space, along with prior mass estimates and prior samples, of size $\hat n_i^\mathscr{S}$ for all slices $i=1,\dots,M$. Although the function $\phi$ is used to generate the partition, the resulting partition can be used to perform stratified Monte Carlo integration on a different $\pi$-integrable function $L:\Omega\to\mathbb{R}$. Using the partition, we can rewrite the integral
\begin{align}\label{eq:q:mean}
\springeronly{\begin{split}}
    \mathcal{Z} &= \int L(\theta)\, \mathrm{d}\pi(\theta) \springerline = \sum_{i=1}^M S_i \int L(\theta) \,\mathrm{d}\pi\vert_{\mathscr{S}_i}(\theta) \springerline = \sum_{i=1}^M S_i\, \cmean{\pi}{\Theta\!\in\!\mathscr{S}_i}{L(\Theta)}\,,
\springeronly{\end{split}}
\end{align}
with the conditional mean defined as
\begin{equation}
    \mathcal{Z}_{\mathscr{S}_i} = \cmean{\pi}{\Theta\!\in\!\mathscr{S}_i}{L(\Theta)}
    = \int L(\theta) \,\mathrm{d}\pi\vert_{\mathscr{S}_i}(\theta)
\end{equation}

We can use the samples within each slice generated by the slicing algorithm to estimate the integral over each slice \textit{via} Monte Carlo integration. This motivates the following estimator, using slice volume estimates $\widehat{S}_i$ and prior samples $\{\Theta^{\mathscr{S}}_{i,k}\}$
\begin{equation}\label{eq:aux1}
    \widehat{\mathcal{Z}} := \sum_{i=1}^M \widehat{S}_i \widehat{\mathcal{Z}}_{\mathscr{S}_i}\quad\text{with}\quad \widehat{\mathcal{Z}}_{\mathscr{S}_i} := \frac{1}{n_i}\sum_{k=1}^{n_i} L(\Theta^{\mathscr{S}}_{i,k})\,,
\end{equation}
where we only selected a subset of points available in each slice. 
If a fixed fraction $\alpha_i\in(0,1]$ of the available points is evaluated in each slice, then
\begin{equation}\label{eq:n_i fixed fraction}
n_i =
\begin{cases}
\alpha_i N_i(1-\hat t_i) = \alpha_i \hat n_i^\mathscr{S}, & i < M,\\
\alpha_M N_{M-1}\hat t_{M-1} = \alpha_M \hat n_M^\mathscr{S}, & i = M.
\end{cases}
\end{equation}
In any case, we assume that at least one point falls within each slice, and that at least one point is evaluated, that is $1 \leq n_i \leq \hat n_i^\mathscr{S}$.

Especially if we assume that the slicing is performed with a cheap surrogate $\phi(\theta)$ of the target function $L(\theta)$, the advantage of this procedure is threefold:

a)~The slicing and the determination of the corresponding constrained prior masses are based on the surrogate, for which it is easy to produce large sample sizes $N_i$, which are necessary for small uncertainties in the prior mass estimates. 
b)~Generating effectively uncorrelated samples from the constrained prior may require several (if not many) Monte Carlo steps involving evaluations of the slicing function. In S$\ell$MC, only the cheap surrogate is evaluated rather than the target in these steps, allowing more sampling effort and hence better decorrelation without additional target-evaluation cost~\citep[for an analysis of sampler effects, see][]{salomone2025smc}.
c)~Once the slices and the corresponding prior masses are available with sufficient accuracy, we can estimate the integral over the slices with much fewer sample points, i.e. the accuracy of the slice volume and the evaluated points of the target function are decoupled. The necessary number of point evaluations per slice depends on the variance 
of $L(\theta)$ within the slice, weighted by the slice's prior mass $S_i$. This can pose a significant advantage over standard reweighting methods, where volume uncertainty and the number of target evaluations are still coupled, as discussed in Section~\ref{sec:RW}.

\subsubsection{Unbiasedness}\label{sec:slmc unbiased}
The estimator $\widehat{\mathcal{Z}}$ is unbiased, i.e. $\mathbb{E}[\widehat{\mathcal{Z}}] = \mathcal{Z}$. 
This follows from two observations: First, the estimator $\hat t_{i}$ of the compression factor $t_i$ is unbiased. Then
\begin{equation}
    \widehat{S}_{i} = (1-\hat t_i)\prod_{k=1}^{i-1} \hat t_k 
\end{equation}
is unbiased for all $i$, since $\hat t_1, \ldots, \hat t_{M}$ are independent. 

Second, the Monte Carlo estimator $\widehat{\mathcal{Z}}_{\mathscr{S}_i}$ is conditionally unbiased for the slice expectation $\mathbb{E}_{\pi}[L(\Theta) \vert \Theta\!\in\!\mathscr{S}_i]$. 

This holds when conditioned on the sigma-field generated by the slice-membership indicators of all sampled points,
\[
\springeronly{\begin{aligned}&}
\mathcal{F}_{\mathscr{S}}
:= \sigma\!\arxivonly{\left(}\springeronly{\bigl(}
\mathbb{1}_{\mathscr{S}_i}(\Theta^{\mathscr{X}}_{i-1,k})
\springerline:\; i=1,\dots,M-1,\; k=1,\dots,N_i
\arxivonly{\right)}\springeronly{\bigr)},
\springeronly{\end{aligned}}
\]
i.e., when knowing which points fall into which sets.

Since the number $n_i$ of selected points within slice $\mathscr{S}_i$ and the corresponding volume estimator $\widehat{S}_i$ are $\mathcal{F}_{\mathscr{S}}$-measurable, they are constant under this conditioning. So we have
\begin{align}
\begin{split}\springeronly{&}
    \mathbb{E}[\widehat{\mathcal{Z}}_{\mathscr{S}_i}\! \vert \mathcal{F}_{\mathscr{S}}] \springerline =\! \frac{1}{n_i}\sum_{k=1}^{n_i} \mathbb{E}[L(\Theta^{\mathscr{S}}_{i,k}) \vert \mathcal{F}_{\mathscr{S}}] \springerline = \mathbb{E}_{\pi}[L(\Theta) \vert\Theta\!\in\!\mathscr{S}_i]
\end{split}
\end{align}
Together this yields
\begin{align}
\begin{split}
    \mathbb{E}[\widehat{\mathcal{Z}}] 
    &=\sum_{i=1}^M \mathbb{E}\!\left[ \mathbb{E}\!\!\left[ \widehat{S}_i\widehat{\mathcal{Z}}_{\mathscr{S}_i} \vert \mathcal{F}_{\mathscr{S}}\right]\!\right] \springerline = \sum_{i=1}^M \mathbb{E}\!\left[ \widehat{S}_i\, \mathbb{E}\!\!\left[  \widehat{\mathcal{Z}}_{\mathscr{S}_i} \vert\mathcal{F}_{\mathscr{S}}\right]\!\right] \\
    &= \sum_{i=1}^M \mathbb{E}[\widehat{S}_i]\, \mathbb{E}_{\pi}[L(\Theta) \vert\Theta\!\in\!\mathscr{S}_i] \springerline = \mathbb{E}_{\pi}[L(\Theta)] = \mathcal{Z}
\end{split}
\end{align}

\subsubsection{Variance}\label{sec:slmc var}
In addition to unbiasedness, we are interested in the variance of the estimator $\widehat{\mathcal{Z}}$. A detailed derivation is provided in Appendix~\ref{app:var slim}; here we present the general structure and discuss its implications. The variance of $\widehat{\mathcal{Z}}$ can be expressed as
\begin{align}
\begin{split}\springeronly{&}
    \mathrm{\mathbb{V}ar}\![\widehat{\mathcal{Z}}] 
    \springerline = \sum_{i=1}^M \!\Bigg(\! \Delta\!\mathcal{Z}_{\mathscr{S}_i}^2 \, \mathbb{E}\left[\frac{\widehat{S}^2_{i}}{n_i}\right] + \mathcal{Z}_{\mathscr{S}_i}^2 \, \mathrm{\mathbb{V}ar}[\widehat{S}_i]
    \springerline + 2\mathrm{\mathbb{C}ov}[\widehat{S}_i,\!\widehat{X}_i] \, \mathcal{Z}_{\mathscr{S}_i} \, \mathcal{Z}_{\mathscr{X}_i} \! \Bigg)
\end{split}
\end{align}
where $n_i \leq \hat n_i^\mathscr{S}$ denotes the number of points selected in each slice for evaluation, with $\hat n_i^\mathscr{S}$ defined in Equation~\eqref{eq:n S i}, and 
\begin{align}
\begin{split}\springeronly{&}
    \mathcal{Z}_{\mathcal{A}} = \mathbb{E}_\pi[L (\Theta)\vert \Theta\!\in\!\mathcal{A}] \springerline\;\;\text{and}\;\;
    \Delta \mathcal{Z}^2_{\mathcal{A}} = \mathrm{\mathbb{V}ar}_\pi[L(\Theta) \vert \Theta\!\in\!\mathcal{A}]\,,
\end{split}
\end{align}
denote the conditional expectation and variance over a non-empty region $\mathcal{A}\in\{\mathscr{X}_i\}_{i=0}^{M-1}\cup\{\mathscr{S}_i\}_{i=1}^M$.
For $i=M$, the covariance contribution is interpreted as zero because $\widehat{X}_M=0$ by convention; no conditional mean over $\mathscr{X}_M$ is required.

This variance decomposition closely mirrors the structure of the nested sampling variance in Equation~\eqref{eq:NS var} that we provided in Section~\ref{sec:NS}, but with key distinctions. In S$\ell$MC, discrete likelihood values are replaced by conditional expectations of the integrand within slices, and an additional term appears that accounts for intra-slice variability. While all terms depend on generally unknown properties of the integrand, they can be estimated from the available samples.

A central advantage of S$\ell$MC compared to standard reweighting techniques is the decoupling of volume estimation from function evaluation. The variance contribution from volume uncertainty, captured in the second and third terms, can be made arbitrarily small by increasing the number of samples $N_i$, without incurring additional evaluations of the target function~$L$. In contrast, the first term, arising from the variance of $L$ within each slice, can be reduced by evaluating more points per slice. If $n_i$ is constant, it can just be pulled out of the expectation. On the other hand, if we choose to evaluate a deterministic fraction $\alpha_i\in(0,1]$ of the points available per slice, using Equation~\eqref{eq:n_i fixed fraction} gives
\begin{equation}
    \mathbb{E}\left[\frac{\widehat{S}_i^2}{n_i}\right] =
    \begin{cases}
    \dfrac{1}{\alpha_i N_i} \dfrac{S_i}{X_{i-1}} \mathbb{E}[\widehat{X}_{i-1}^2], & i < M,\\[8pt]
    \dfrac{1}{\alpha_M N_{M-1}} \dfrac{S_M}{X_{M-2}} \mathbb{E}[\widehat{X}_{M-2}^2], & i = M.
    \end{cases}
\end{equation}

This flexibility enables efficient resource allocation: volume estimation can be refined independently of the function evaluation budget. S$\ell$MC is therefore especially well suited for problems where the integrand exhibits low variance within each slice, as high-accuracy integration can be achieved with minimal function evaluations. To achieve this, the slicing function $\phi$ should ideally act as a surrogate for the integrand $L$, such that $L$ exhibits low variability within each slice.

\subsubsection{Redrawing Points for Given Thresholds}\label{sec:combining runs}
In practice, it is often beneficial to be able to redraw points for given thresholds, to increase the accuracy of the volume estimates, or to generate more points within each slice. Different slicing runs, which have the same thresholds, can be easily combined. Each slicing run $r$ with thresholds $\{\phi_j\}$, has samples $\{\Theta_{j,k}^{\mathscr{X},(r)}\}$ within the level sets $\{\mathscr{X}_j^{(r)}\}$. Assuming that the samples are independent for each run, we can simply pool them at each step of the algorithm
\begin{equation}
    \springeronly{\begin{aligned}&}
    \{\Theta^{\mathscr{X}}_{i,k}\} = \bigcup_{r} \{\Theta_{i,k}^{\mathscr{X},(r)}\}\springerline\quad\text{and}\quad \{\Theta^{\mathscr{S}}_{i,k}\} = \bigcup_{r} \{\Theta_{i,k}^{\mathscr{S},(r)}\}\,,
    \springeronly{\end{aligned}}
\end{equation}
and thereby increase the effective sample size to $N_i = \sum_r N_i^{(r)}$, and enhance the estimate of the updated compression factors
\begin{equation}
    \hat{t}_i = \frac{\hat n_i^\mathscr{X}}{N_i}\,\quad\text{with}\quad \hat n_i^\mathscr{X} = \sum_r \hat n_i^{\mathscr{X},(r)}\,.
\end{equation}
The rest of the analysis remains the same. An adaptation of this procedure for the case of different thresholds is provided in Appendix~\ref{app:redrawing}.

\subsubsection{Posterior Expectations and Posterior Samples}
If the target function $L$ is non-negative, we may interpret it as a likelihood function. Then the integral $\mathcal{Z}$ is the evidence. 
In this case, it is also possible to compute the expectation value of a function $g:\Omega\to\mathbb{R}$ 
with respect to the posterior distribution $\mathrm{d}p(\theta)=L(\theta)\,\mathrm{d}\pi(\theta) / \mathcal{Z}$
\begin{align}
\begin{split}\springeronly{&}
    \mathcal{G} = \int_\Omega g(\theta ) \, 
    \frac{L(\theta)}{\mathcal{Z}} \,\mathrm{d}\pi(\theta) \springerline =   \sum_{i=1}^M S_i\int g(\theta ) \,
    \frac{L(\theta)}{\mathcal{Z}} \,\mathrm{d}\pi\vert_{\mathscr{S}_i}(\theta)
\end{split}
\end{align}
using S$\ell$MC with the previously determined sample $\{\Theta^{\mathscr{S}}_{i,k} \}$ of the constraint prior
\begin{align}
\begin{split}
    \widehat{\mathcal{G}} = \sum_{i=1}^M \frac{\widehat{S}_i}{n_i} \sum_{k=1}^{n_i} g(\Theta^{\mathscr{S}}_{i,k} ) \, 
    \frac{L(\Theta^{\mathscr{S}}_{i,k})}{\widehat{\mathcal{Z}}} 
\end{split}
\end{align}
Similar to NS, the slicing procedure naturally induces weighted posterior representatives
\begin{equation}
    p_{i,k} := \frac{\widehat{S}_i L(\Theta_{i,k}^{\mathscr{S}})}{n_i\, \widehat{\mathcal{Z}}}\,,
\end{equation}
which can be used to express the posterior expectation differently
\begin{align}
\begin{split}
    \widehat{\mathcal{G}} = \sum_{i=1}^M \sum_{k=1}^{n_i} g(\Theta^{\mathscr{S}}_{i,k} ) \,  p_{i,k}\,.
\end{split}
\end{align}
Here $p_{i,k}$ is the probability for a parameter value $\theta$ in a small volume of size $\Delta V_{i}$ centered at 
$\Theta^{\mathscr{S}}_{i,k}$, with $\Delta V_{i} \approx S_i/n_i$. Note that the number of points $n_i$ evaluated for $g$ and $L$ in slice $i$ can be adjusted depending on the variability of $g(\theta)\cdot p(\theta)$.

Beyond weighted representatives, S$\ell$MC also enables the generation of equally weighted samples from the posterior via approximate rejection sampling. For that purpose, we define the piecewise-constant envelope
\begin{equation}
    L^\mathscr{S} := \sum_{i=1}^M L_{\mathscr{S}_i}^{\max} \cdot \mathbb{1}_{\mathscr{S}_i}, \quad
    L_{\mathscr{S}_i}^{\max} := \max_{\theta \in \mathscr{S}_i} L(\theta)\,.
\end{equation}
As $L_{\mathscr{S}_i}^{\max}$ is typically unknown, we have to estimate it as $\hat{L}_{\mathscr{S}_i}^{\max} := \max \{L(\Theta_{i,k}^{\mathscr{S}})\}$ to obtain an estimated envelope $\hat{L}^{\mathscr{S}}$.

To sample from the posterior, we first draw from the approximate (unnormalized) distribution $\hat{L}^\mathscr{S}$ with prior distribution $\pi$ using the estimated volumes $\widehat{S}_i$ (see Appendix~\ref{app:draw simple}), and then apply rejection sampling within the selected slice $\mathscr{S}_i$. The acceptance probability is proportional to $L(\theta)/L_{\mathscr{S}_i}^{\max}$, from which it can be seen that the efficiency of this step depends on the within-slice variability of $L$.

S$\ell$MC thus combines the variance-reduction benefits of stratification with the computational efficiency of surrogate-assisted reweighting, and provides a principled framework for surrogate-informed slice construction and target-evaluation budgeting, while additionally enabling the generation of posterior samples.

\section{Numerical analysis}\label{sec:experiments}

To assess the performance of S$\ell$MC, we compare it with surrogate-based reweighting (RW), a Nested-Sampling reference benchmark (NS), and Monte Carlo importance sampling (IS) on controlled Gaussian problems in which the mismatch between a surrogate and a target can be varied continuously.
For S$\ell$MC and RW, the surrogate determines the ordering and constrained regions. The NS curve shown below is a reference calculation performed on the isotropic surrogate Gaussian and reused at every target-deformation value. It is therefore not a separate direct-NS run on every shifted or anisotropic target. We label it ``NS (surrogate benchmark)'' throughout.
For NS and RW, we use the deterministic expected compression $X_i=[N_{\mathrm{live}}/(N_{\mathrm{live}}+1)]^i$ and a quadrature based on the discarded representatives. Details on the numerical setup for the respective methods are provided in Appendix~\ref{app:numerical_implementation}.
The goal is to compare how the different estimators translate surrogate mismatch into the required number of target evaluations and assess the robustness of the respective methods to surrogate-target mismatch.
For this controlled comparison, we use independent exact constrained draws to isolate the effect of surrogate--target mismatch. In S$\ell$MC, sampler effects can be reduced through additional surrogate-only sampling without increasing the target-evaluation cost; their influence in the likelihood-driven setting without a separate surrogate is investigated by \citet{salomone2025smc}.

\subsection{Benchmark models}\label{sec:models}

The benchmark models are Gaussian surrogate-target pairs, for which the mismatch can be varied continuously through explicit parameters while the remaining problem setup is held fixed. This permits changes in performance of the estimator, which can be attributed directly to surrogate mismatch rather than to application-specific model structure.
In the implementation, $\Omega=B_{0.5}(\mu_\phi)\subset\mathbb{R}^d$ with $\mu_\phi=(0.5,\ldots,0.5)$, and the base measure is uniform volume on this ball. The Gaussian parameters are chosen such that both $\phi$ and $L$ are well contained in $\Omega$, so boundary effects are negligible and the reference value is given by the corresponding full-space Gaussian integral, which can be computed in closed form. 
Let
\begin{equation}
\varphi_d(\theta;\mu,\Sigma)
:=
\exp\!\left[-\frac{1}{2}(\theta-\mu)^\top\Sigma^{-1}(\theta-\mu)\right]
\end{equation}
denote an unnormalized Gaussian density on $\mathbb{R}^d$. For simplicity we choose an isotropic Gaussian as surrogate
\begin{equation}
\phi(\theta)=\varphi_d(\theta;\mu_\phi,\sigma_\phi^2 I_d)\,,
\end{equation}
with variance $\sigma_\phi^2=2.5\times10^{-5}$ (standard deviation $\sigma_\phi=0.005$).

We consider two target families -- shift and anisotropy.
The shift benchmark varies the relative mean displacement while keeping the variances isotropic with $\sigma_L^2=\sigma_\phi^2$,
\begin{equation}
\springeronly{\begin{aligned}&}
L_{\mathrm{shift}}(\theta;b)
=\varphi_d(\theta;\mu_L,\sigma_L^2 I_d)\,,
\springerline\qquad
b=\frac{\|\mu_L-\mu_\phi\|}{\sigma_L}\,.
\springeronly{\end{aligned}}
\end{equation}
The anisotropy benchmark fixes the means $\mu_L=\mu_\phi$ and varies
\begin{align}
\begin{split}
L_{\mathrm{aniso}}(\theta;a)
&=\varphi_d(\theta;\mu_L,\Sigma_L(a))\,,\\
\Sigma_L(a)
&=\sigma_\phi^2\,\,\mathrm{diag}(\underbrace{a^2,\ldots,a^2}_{\lfloor d/2\rfloor},
\underbrace{a^{-2},\ldots,a^{-2}}_{\lceil d/2\rceil})\,,
\springerline\quad a\ge 1\,.
\end{split}
\end{align}
Thus, one half of the coordinate directions is stretched and the other half is compressed.
Note that an isotropic mismatch would only influence the performance of IS, while the other methods are invariant to isotropic scaling of the surrogate.
For S$\ell$MC, this invariance reflects that the surrogate enters the slice construction through its induced ordering rather than through pointwise agreement with the target.

Thus the shift family probes mean mismatch, while the anisotropy family probes covariance mismatch. For the even dimensions used below, the product of all target widths in the anisotropy benchmark is independent of $a$. For each mismatch setting, the exact value $\mathcal{Z}$ is used to report relative errors $
\delta := |\widehat{\mathcal{Z}} \!-\! \mathcal{Z}| / \mathcal{Z}$.

\subsection{Benchmarking procedure}

We compare the above methods by evaluating their relative error as a function of the target-function evaluation cost
\begin{equation}
Q:=\#\{\text{required evaluations of }L\}\,.
\end{equation}
Evaluations of $\phi$ are treated as negligible and are not counted in $Q$.
For a fixed cost $Q$, the relative error $\delta(Q)$ is random because each method is stochastic.
Writing $\mathbb{P}[\,\cdot\,]$ for the probability distribution at fixed $Q$ for the different methods, we define the minimal required cost for a tolerance $\epsilon$ and success probability $p$ as
\begin{equation}
Q^\star(\epsilon,p)=\min\left\{Q:\;\mathbb{P}\!\left[\delta(Q)\le\epsilon\right]\ge p\right\}\,,
\end{equation}
which is the minimal number of point evaluations of the target function required to achieve a relative error of at most $\epsilon$ with probability at least $p$.
For S$\ell$MC, RW, and the NS surrogate benchmark, every scheduled budget is executed as an independent calculation. The schedules are defined in floating point, and every occurrence of the resulting integer budget is retained. Thus, occurrences that round to the same integer budget are independent repetitions, not duplicated data or checkpoints from one progressive trajectory.
We estimate $Q^\star$ from the empirical pairs $(Q_j,\delta_j)$ with an inclusive centred multiplicative window $[Q_j/1.5,1.5Q_j]$. A candidate qualifies when at least a fraction $p=0.9$ of all observations in this interval satisfy $\delta\le\epsilon$. If the lower edge lies below the minimum simulated cost, the available observations are used; complete upper coverage is required. Uncertainty is quantified with 2,000 bootstrap resamples. We report the median and 16th/84th percentiles, and display a result only when at least 80\% of the bootstrap resamples produce a qualifying window. Details are given in Appendix~\ref{app:numerical_implementation}.
For IS, $Q^\star$ is obtained analytically using the central limit theorem; see Appendix~\ref{app:IS budget}.

\begin{figure}[t]
  \centering
  \includegraphics[width=0.94\linewidth]{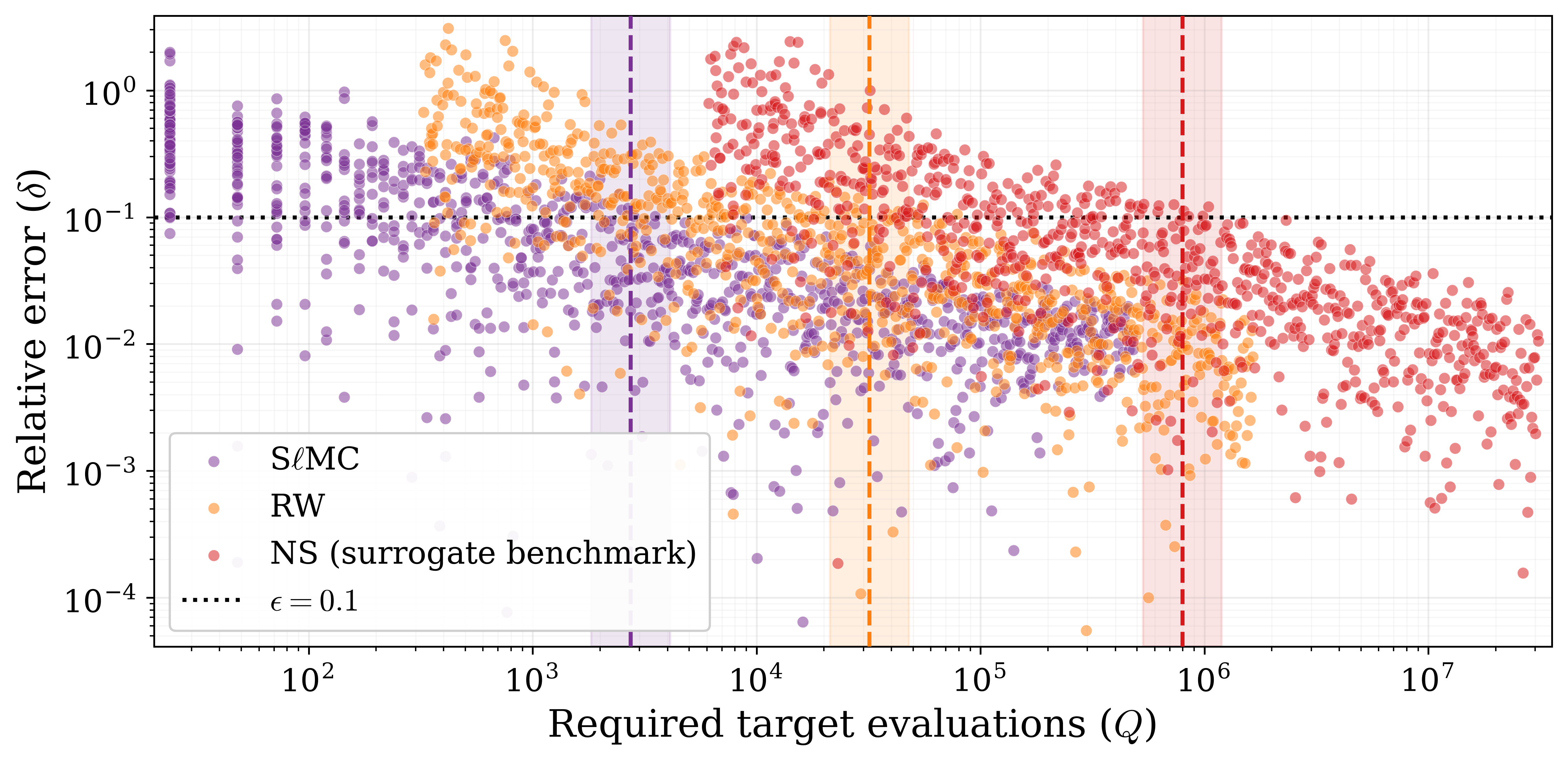}
  \caption{Relative error $\delta$ as a function of target-evaluation cost $Q$ for the anisotropy benchmark at $a=2.4$ in $d=2$, using the final independent-occurrence schedules for S$\ell$MC (purple), RW (orange), and the NS surrogate benchmark (red). Repeated integer budgets remain separate observations. Transparent bands indicate the first qualifying centred windows $[Q^\star/1.5,1.5Q^\star]$ for $\epsilon=0.1$ and $p=0.9$; available lower-bound observations are used and complete upper coverage is required. Dashed vertical lines mark the corresponding direct estimates of $Q^\star$ (see Appendix~\ref{app:numerical_implementation}).}
  \label{fig:results_scatter_windows_anisotropy_d2}
\end{figure}

\subsection{Results}

\begin{figure*}[tb]
  \centering
  \includegraphics[width=\textwidth]{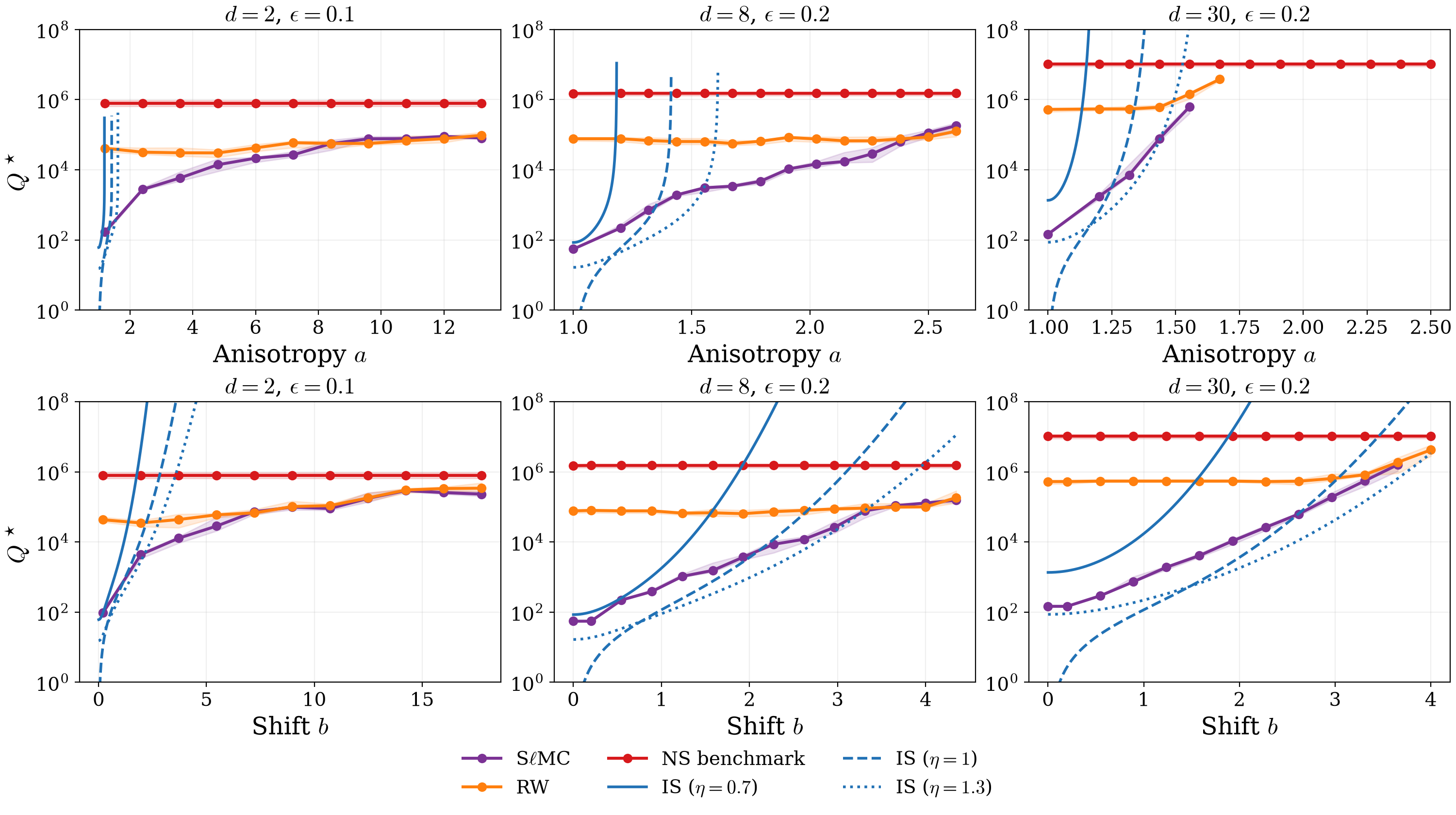}
  \caption{Required target-evaluation cost $Q^\star$ across benchmark families and dimensions.
  The top row shows the anisotropy benchmark and the bottom row the shift benchmark, with columns corresponding to $d=2$, $d=8$, and $d=30$.
  The $d=2$ column uses $\epsilon=0.1$, while the $d=8$ and $d=30$ columns use $\epsilon=0.2$; the latter two include the exact-agreement points $a=1$ and $b=0$. Empirical results require at least 90\% success in the centred window $[Q^\star/1.5,1.5Q^\star]$. Available lower-bound observations are used and complete upper coverage is required. Shaded bands show bootstrap 16th/84th percentiles from 2,000 resamples; results with less than 80\% bootstrap support are omitted. The NS curve is the reused surrogate-Gaussian benchmark. IS curves are analytical CLT estimates for proposal variances $\eta\sigma_\phi^2$ with $\eta\in\{0.7,1.0,1.3\}$.}
  \label{fig:results_nstar_dims}
\end{figure*}

We report benchmark results for $d=2$, $d=8$, and $d=30$.
For clarity, the $d=2$ panels use $\epsilon=0.1$, while the $d=8$ and $d=30$ panels use $\epsilon=0.2$.
All panels in Figure~\ref{fig:results_nstar_dims} show the required cost $Q^\star(\epsilon, p=0.9)$, with bootstrap 16th/84th percentile bands for S$\ell$MC, RW, and the NS surrogate benchmark (see Appendix~\ref{app:numerical_implementation}), and analytical CLT values for IS (see Appendix~\ref{app:IS budget}).
For IS we additionally show the effect of scaling the surrogate proposal variance by using
\begin{equation}
\phi_\eta(\theta)
=\varphi_d(\theta;\mu_\phi,\eta\sigma_\phi^2 I_d)\,,
\qquad \eta\in\{0.7,1.0,1.3\}\,.
\end{equation}
The other methods are invariant to isotropic scaling of the surrogate, so only one curve is shown for each of S$\ell$MC, RW, and NS.

The NS reference curve is constant because the same isotropic surrogate-Gaussian benchmark is reused at every target-deformation value. Its larger cost compared with RW mainly reflects the constrained-replacement overhead $c_{\mathrm{NS}}$ in the cost convention of Appendix~\ref{app:numerical_costs}. 
In the anisotropy benchmark, the top-row panels of Figure~\ref{fig:results_nstar_dims} show $Q^\star$ as a function of the anisotropy parameter~$a$.
Across all dimensions, IS has the lowest cost when $a$ is close to one, as expected from its favorable behavior when the proposal is nearly identical to the target.
This advantage is fragile: as soon as the surrogate is too narrow in one target direction, the importance-weight variance diverges.
For the unscaled proposal, the finite-variance condition derived in Appendix~\ref{app:IS budget} translates to $a<\sqrt{2}$ for the stretched directions; with proposal variance scale $\eta$, this boundary becomes $a<\sqrt{2\eta}$.
Thus the IS curves terminate at predictable finite-variance boundaries, and the deterioration before those boundaries is particularly pronounced.

S$\ell$MC and RW are more robust because they do not rely on a single global importance density with finite weight variance.
Instead, both methods use the surrogate to partition the parameter space and distribute samples across the induced level sets, so their cost grows more gracefully as the surrogate-target mismatch increases.
Both methods become more expensive as $a$ increases, since the target function varies more strongly inside the surrogate-induced slices. S$\ell$MC generally remains below RW at small and moderate mismatch, while the curves approach and can cross at the largest supported anisotropies.
This is the practical effect of decoupling slice-volume estimation from target evaluation: S$\ell$MC can spend many cheap surrogate samples on the partition and volume estimation, while only the selected integration points require expensive evaluations of $L$.
As the mismatch and thus the within-slice variance increase S$\ell$MC gets closer to RW, as many points within each slice are needed to obtain a small relative error.
This behavior is consistent across all investigated dimensions. However, for increasing dimension the mismatch leads to higher absolute $Q^\star$ values and more sensitive behavior with respect to the mismatch.

In the shift benchmark, the bottom-row panels of Figure~\ref{fig:results_nstar_dims} show $Q^\star$ as a function of the shift parameter $b$.
Here the surrogate and target have the same shape, but their centers are displaced.
IS again starts with the lowest cost near perfect agreement; for the unscaled proposal its normalized variance grows like $\exp(b^2)$ in the isotropic shift case (Appendix~\ref{app:IS budget}), and the variance-scaled proposals retain the same exponential sensitivity.
This is less abrupt than the anisotropy benchmark because there is no finite-variance boundary; nevertheless, the cost eventually becomes enormous as the high-mass regions of surrogate and target separate.

S$\ell$MC and RW are more stable under this mismatch because their surrogate-based partitions still place samples across the relevant level sets.
Their costs increase with $b$, but the growth is much slower than for IS over the range shown. S$\ell$MC is generally below RW at small and moderate shifts, while the methods become comparable at the largest shifts.
This is consistent with the same decoupling principle: the surrogate partition can still be estimated cheaply, while target evaluations are allocated only to estimate the contributions within the resulting slices.

The two benchmark families highlight complementary failure modes.
IS is optimal only when the surrogate is very accurate: it is extremely sensitive to shape mismatch, where a single overly narrow surrogate direction can make the variance infinite, and it deteriorates exponentially under pure position mismatch.
RW and S$\ell$MC are more robust because they use the surrogate to partition the parameter space rather than using a single global importance density.

S$\ell$MC is typically more efficient than RW because the expensive target evaluations are decoupled from the surrogate-based volume estimation, but this advantage naturally shrinks when large mismatch forces many target evaluations within each slice.
This also explains why the two curves approach each other at larger mismatch.
When the target varies strongly within the surrogate-induced slices, S$\ell$MC must evaluate many target points per slice, so within-slice target variance dominates and slice-volume uncertainty becomes less relevant.
In the same regime, the volume uncertainty in RW becomes small and the remaining uncertainty stems from the target variance along the surrogate level sets. So S$\ell$MC and RW have the same dominant source of uncertainty.
Additionally, the effective point weights of RW and S$\ell$MC become similar, when RW and S$\ell$MC use a comparable number of points per volume.
For small mismatch, in contrast, S$\ell$MC can achieve low error with few target evaluations per slice due to the decoupling of volume uncertainty and target evaluations, while RW still pays for the coupled volume-estimation and target-evaluation mechanism.

Some $d=30$ curves terminate before the end of the deformation range. These points were simulated, but the available scan did not yield a qualifying convergence window in at least 80\% of the bootstrap resamples; unsupported results are therefore omitted.

The uncertainty shown here is conditional on one fixed surrogate-slice cache in each dimension. Independent occurrences resample points from that immutable cache but do not regenerate the slice geometry or its volume estimates.

Appendix~\ref{app:slmc_variants} complements this comparison with a control-variate formulation that combines an accurately determined surrogate integral with slice-wise estimates of a scaled target-surrogate residual.

Together, the results show that S$\ell$MC turns surrogate quality into reduced target-evaluation cost while retaining considerably more robustness to mismatch than direct importance sampling.

\section{Conclusion and Outlook}\label{sec:conclusion}

We have introduced Slice Monte Carlo integration (S$\ell$MC) as a surrogate-informed integration method that uses a Nested Sampling-like construction to partition the parameter space into slices adapted to the surrogate level sets. The partition and the slice volumes are estimated using the cheap surrogate, while the expensive target function is evaluated only for the Monte Carlo estimates within the slices.

The central advantage is the decoupling of slice-volume estimation from target evaluation. Standard reweighting with Nested Sampling uses the surrogate to avoid expensive constrained sampling, but target evaluations remain tied to the quadrature resolution. S$\ell$MC separates these two tasks: many surrogate samples can refine the slice volumes, while the number of target evaluations can be chosen independently according to the integration budget. This additional degree of freedom explains the reduced target cost relative to reweighting at small and moderate mismatch. At strong mismatch, the within-slice target variance dominates and the cost of S$\ell$MC approaches or can exceed that of reweighting.

Compared with classical importance sampling, S$\ell$MC is substantially more robust to surrogate-target mismatch. Importance sampling is most efficient when the surrogate proposal is already very close to the target, but it deteriorates rapidly when surrogate-target mismatch occurs. S$\ell$MC instead uses the surrogate only to define a partition of the parameter space and estimates the target contribution within each slice directly, avoiding the severe failure modes of importance sampling. The price is that, in the ideal case of nearly perfect surrogate-target agreement, direct importance sampling can require fewer target evaluations. In realistic applications, however, surrogate information is rarely perfect, and robustness to mismatch is often more valuable than optimal performance in this limiting case.

The primary numerical comparison uses the transparent uniform allocation $n_i=n$ for all slices $i=1,\ldots,M$. A possible direction for future work is to extend S$\ell$MC to multifidelity or multilevel formulations, where richer surrogate information may justify additional complexity~\citep{peherstorfer2016mfmc, peherstorfer2018mfmc, Giles2015mlmc}.

S$\ell$MC is particularly attractive because the hard constrained-sampling task is performed on the cheap surrogate, while the resulting slices are then used for a stratified integration of the expensive target. This makes the method relevant for applications where low-fidelity physical models are readily available~\citep{najm2009, Yun2016uq}, where simplified energy landscapes can guide expensive partition-function calculations~\citep{partay2021, binder2022ridge}, or where machine-learned surrogate models approximate costly high-fidelity evaluations~\citep{Lu2019surr, krondorfer2023gpr, kovacs2025maceoff, Lu2022}. Overall, S$\ell$MC provides a robust and flexible framework for surrogate-informed Bayesian evidence integration when the surrogate is useful but not perfectly reliable.

\begin{appendix}

\section{Variance of Slice Integration}\label{app:var slim}
The aim of this section is the computation of the variance of the S$\ell$MC estimator of the evidence $\mathcal Z.$ Recall that the estimator of the constrained prior mass is given by $\widehat{X}_{i-1} = \prod_{k=1}^{i-1} \hat t_k$ with independent compression factors $N_k \hat t_k \sim \text{Binomial}(N_k, X_k / X_{k-1})$ for $k<M$, with $\widehat{X}_0 = X_0 = 1$ and $\widehat{X}_M=X_M=0$ by convention. The estimator of the slice volume is $\widehat{S}_k = \widehat{X}_{k-1} - \widehat{X}_k$. From that, we obtain
\begin{align}
\begin{split}
    \mathbb{E}[\widehat{X}_{i-1}] 
    &= \prod_{k=1}^{i-1} \mathbb{E}[\hat t_k] = \prod_{k=1}^{i-1} \frac{X_k}{X_{k-1}} = X_{i-1} \\
    \mathbb{E}[\widehat{X}_{i-1}^2]
    &= X_{i-1}^2 + \mathrm{\mathbb{V}ar}[\widehat{X}_{i-1}] \\
\end{split}
\end{align}
with variance
\begin{align}
\begin{split}
    \mathrm{\mathbb{V}ar}[\widehat{X}_{i-1}]
    &= \prod_{k=1}^{i-1}\!\left(\!\frac{1}{N_k}\frac{X_k}{X_{k-1}} \! \left( 1 \!-\! \frac{X_k}{X_{k-1}}\! \right) + \!\frac{X_k^2}{X_{k-1}^2} \! \right) 
    \springerline - \prod_{k=1}^{i-1} \frac{X^2_k}{X^2_{k-1}} \\
    &= X_{i-1}^2\sum_{k=1}^{i-1} \frac{S_k}{N_kX_k} + \mathcal{O}\!\left(\!\frac{1}{N^2}\!\right)\,.
\end{split}
\end{align}
For $i<M$, we can compute the variance of the slice estimator $\widehat{S}_i$ by using the independence of $\widehat{X}_{i-1}$ and $\hat t_i$
\begin{align}
\begin{split}
    \springeronly{&}\mathrm{\mathbb{V}ar}[\widehat{S}_i] \springeronly{\\}
    &= \mathrm{\mathbb{V}ar}[\widehat{X}_{i-1} (1\!-\!\hat t_i)] \\
    &= \mathrm{\mathbb{V}ar}[\widehat{X}_{i-1}]\mathrm{\mathbb{V}ar}[1\!-\!\hat t_i] \springerline + \mathrm{\mathbb{V}ar}[\widehat{X}_{i-1}]\mathbb{E}[1\!-\!\hat t_i]^2
    \springerline + \mathrm{\mathbb{V}ar}[1\!-\!\hat t_i]\mathbb{E}[\widehat{X}_{i-1}]^2 \\
    &= \mathrm{\mathbb{V}ar}[\widehat{X}_{i-1}]\left(\!\frac{1}{N_i}\frac{S_i}{X_{i-1}}\!\left( 1\!-\! \frac{S_i}{X_{i-1}} \right) + \frac{S_i^2}{X_{i-1}^2} \right) \springerline + \frac{S_i X_i}{N_i} \\
    &= S_i^2 \sum_{k=1}^{i-1}\frac{S_k}{N_k X_k} + \frac{S_i X_i}{N_i} + \mathcal{O}\!\left(\!\frac{1}{N^2}\!\right)\,.
\end{split}
\end{align}
For the final slice, $\widehat{S}_M=\widehat{X}_{M-1}$, so $\mathrm{\mathbb{V}ar}[\widehat{S}_M]=\mathrm{\mathbb{V}ar}[\widehat{X}_{M-1}]$.
Another quantity we need for computing $\mathrm{\mathbb{V}ar}[ \widehat{\mathcal Z}]$, is the covariance between the estimators $\widehat{S}_i$ and $\widehat{X}_i$ for $i<M$, which is
\begin{align}
\begin{split}
    \springeronly{&}\mathrm{\mathbb{C}ov}[\widehat{S}_i,\widehat{X}_i] \springeronly{\\}
    &= \mathrm{\mathbb{C}ov}[\widehat{X}_{i-1} (1\!-\!\hat t_i), \widehat{X}_{i-1} \hat t_i] \\
    &= \mathbb{E}[\widehat{X}_{i-1}^2] \mathrm{\mathbb{C}ov}[1\!-\!\hat t_i,\hat t_i] \springerline + \mathrm{\mathbb{V}ar}[\widehat{X}_{i-1}] \mathbb{E}[1\!-\!\hat t_i] \mathbb{E}[\hat t_i] \\
    &= \mathrm{\mathbb{V}ar}[\widehat{X}_{i-1}] \left( \mathbb{E}[\hat t_i] - \mathbb{E}[\hat t_i]^2 \right) \springerline - \mathrm{\mathbb{V}ar}[\hat t_i] \mathbb{E}[\widehat{X}_{i-1}^2] \\
    &= \frac{X_i}{X_{i-1}}\left(\! 1\!-\!\frac{X_i}{X_{i-1}}\! \right)\!\left( \mathrm{\mathbb{V}ar}[\widehat{X}_{i-1}] \!-\! \frac{1}{N_i} \mathbb{E}[\widehat{X}_{i-1}^2] \right) \\
    &= X_i S_i \left( \sum_{k=1}^{i-1} \frac{S_k}{N_k X_k} - \frac{1}{N_i} \right) + \mathcal{O}\!\left(\!\frac{1}{N^2}\!\right)\,.
\end{split}
\end{align}

The previous computations allow for determining the variance of the estimator $\widehat{\mathcal{Z}} = \sum_{i=1}^M \widehat{S}_i \widehat{\mathcal{Z}}_{\mathscr{S}_i}$, yielding
\begin{align}
\begin{split}\springeronly{&}
    \mathrm{\mathbb{V}ar}[\widehat{\mathcal{Z}}] = \sum_{i=1}^M \mathrm{\mathbb{V}ar}[\widehat{S}_i \widehat{\mathcal{Z}}_{\mathscr{S}_i}] \springerline + \! 2\!\!\sum_{\substack{i,j =1 \\ i<j}}^M\!\! \mathrm{\mathbb{C}ov}[\widehat{S}_i \widehat{\mathcal{Z}}_{\mathscr{S}_i}, \widehat{S}_j \widehat{\mathcal{Z}}_{\mathscr{S}_j}]\,.
\end{split}
\end{align}
We again can compute the individual terms by conditioning on the sigma-field $\mathcal{F}_{\mathscr{S}}$, which encodes the slice-membership information of the sampled points. By using the law of total variance and the $\mathcal{F}_{\mathscr{S}}$-measurability of $\widehat{S}_i$ and $n_i$, we get
\begin{align}
\begin{split}
    \mathrm{\mathbb{V}ar}[\widehat{S}_i \widehat{\mathcal{Z}}_{\mathscr{S}_i}] 
    &\!=\! \mathbb{E}[\mathrm{\mathbb{V}ar}[\widehat{S}_i \widehat{\mathcal{Z}}_{\mathscr{S}_i} \vert \mathcal{F}_{\mathscr{S}}]] \springerline + \mathrm{\mathbb{V}ar}[\mathbb{E}[\widehat{S}_i \widehat{\mathcal{Z}}_{\mathscr{S}_i} \vert \mathcal{F}_{\mathscr{S}}]] \\
    &\!=\! \mathrm{\mathbb{V}ar}[L(\Theta)\vert\Theta\!\in\!\mathscr{S}_i]\;\mathbb{E}[\frac{\widehat{S}_i^2}{n_i}] \springerline + \mathbb{E}[L(\Theta)\vert\Theta\!\in\!\mathscr{S}_i]^2\; \mathrm{\mathbb{V}ar}[\widehat{S}_i]\,
\end{split}
\end{align}
and for the covariance
\begin{align}
\begin{split}\springeronly{&}
    \mathrm{\mathbb{C}ov}[\widehat{S}_i \widehat{\mathcal{Z}}_{\mathscr{S}_i}, \widehat{S}_j \widehat{\mathcal{Z}}_{\mathscr{S}_j}] \springeronly{\\} &= \mathbb{E}[\mathrm{\mathbb{C}ov}[\widehat{S}_i \widehat{\mathcal{Z}}_{\mathscr{S}_i}, \widehat{S}_j \widehat{\mathcal{Z}}_{\mathscr{S}_j}\vert \mathcal{F}_{\mathscr{S}}]] \\
    &\quad + \mathrm{\mathbb{C}ov}[\mathbb{E}[\widehat{S}_i \widehat{\mathcal{Z}}_{\mathscr{S}_i}\vert \mathcal{F}_{\mathscr{S}}], \mathbb{E}[\widehat{S}_j \widehat{\mathcal{Z}}_{\mathscr{S}_j}\vert \mathcal{F}_{\mathscr{S}}]]\,.
\end{split}
\end{align}
Using again the $\mathcal{F}_{\mathscr{S}}$-measurability and the independence of the samples within each slice, we get for the first term for $i\neq j$
\begin{align}
\begin{split}\springeronly{&}
    \mathbb{E}[\mathrm{\mathbb{C}ov}[\widehat{S}_i \widehat{\mathcal{Z}}_{\mathscr{S}_i}, \widehat{S}_j \widehat{\mathcal{Z}}_{\mathscr{S}_j}\vert \mathcal{F}_{\mathscr{S}}]] 
    \springerline =\mathbb{E}[\widehat{S}_i \widehat{S}_j \mathrm{\mathbb{C}ov}[ \widehat{\mathcal{Z}}_{\mathscr{S}_i}, \widehat{\mathcal{Z}}_{\mathscr{S}_j}\vert \mathcal{F}_{\mathscr{S}}]] = 0
\end{split}
\end{align}
and thus we get
\begin{align}
\begin{split}
    \mathrm{\mathbb{C}ov}[\widehat{S}_i \widehat{\mathcal{Z}}_{\mathscr{S}_i}, \widehat{S}_j \widehat{\mathcal{Z}}_{\mathscr{S}_j}] = \mathcal{Z}_{\mathscr{S}_i} \mathcal{Z}_{\mathscr{S}_j} \mathrm{\mathbb{C}ov}[\widehat{S}_i,\widehat{S}_j]\,.
\end{split}
\end{align}
For $i < j$ we can write
\begin{equation}
    \widehat{S}_j = \widehat{X}_i \prod_{k=i+1}^{j-1} \hat t_k \;\;(1\!-\!\hat t_j)\,,
\end{equation}
and again, by using the independence of the compression factors, we obtain
\begin{align}
\begin{split}\springeronly{&}
    \mathrm{\mathbb{C}ov}[\widehat{S}_i,\widehat{S}_j] 
    \springerline = \mathrm{\mathbb{C}ov}[\widehat{S}_i,\widehat{X}_i] \mathbb{E}[1\!-\!\hat t_j] \prod_{k=i+1}^{j-1}\mathbb{E}[\hat t_k]
    \springerline = \mathrm{\mathbb{C}ov}[\widehat{S}_i,\widehat{X}_i] \frac{S_j}{X_i}\,.
\end{split}
\end{align}
Evaluating the sum over $j$ in the covariance term yields
\begin{align}
\begin{split}\springeronly{&}
    \sum_{j=i+1}^M \mathrm{\mathbb{C}ov}[\widehat{S}_i \widehat{\mathcal{Z}}_{\mathscr{S}_i}, \widehat{S}_j \widehat{\mathcal{Z}}_{\mathscr{S}_j}] 
    \springerline = \frac{\mathcal{Z}_{\mathscr{S}_i} \mathrm{\mathbb{C}ov}[\widehat{S}_i,\widehat{X}_i]}{X_i} \sum_{j=i+1}^M  \mathcal{Z}_{\mathscr{S}_j}  S_j
    \springerline = \mathcal{Z}_{\mathscr{S}_i} \mathcal{Z}_{\mathscr{X}_i} \mathrm{\mathbb{C}ov}[\widehat{S}_i,\widehat{X}_i]\,.
\end{split}
\end{align}
Altogether, this gives the final expression for the variance of the estimator $\widehat{\mathcal{Z}}$
\begin{align}
\begin{split}\springeronly{&}
    \mathrm{\mathbb{V}ar}\![\widehat{\mathcal{Z}}] 
    \springerline = \sum_{i=1}^M \!\Bigg(\! \Delta\!\mathcal{Z}_{\mathscr{S}_i}^2 \, \mathbb{E}[\frac{\widehat{S}^2_{i}}{n_i}] + \mathcal{Z}_{\mathscr{S}_i}^2 \, \mathrm{\mathbb{V}ar}[\widehat{S}_i]
    \springerline + 2\mathrm{\mathbb{C}ov}[\widehat{S}_i,\!\widehat{X}_i] \, \mathcal{Z}_{\mathscr{S}_i} \, \mathcal{Z}_{\mathscr{X}_i} \! \Bigg)
\end{split}
\end{align}
where $n_i$ denotes the number of points selected in each slice, and
\begin{align}
\begin{split}\springeronly{&}
    \mathcal{Z}_{\mathcal{A}} = \mathbb{E}_\pi[L (\Theta)\vert \Theta\!\in\!\mathcal{A}] \springerline\;\;\text{and}\;\;
    \Delta \mathcal{Z}^2_{\mathcal{A}} = \mathrm{\mathbb{V}ar}_\pi[L(\Theta) \vert \Theta\!\in\!\mathcal{A}]\,,
\end{split}
\end{align}
denote the conditional expectation and variance over a non-empty region $\mathcal{A}\in\{\mathscr{X}_i\}_{i=0}^{M-1}\cup\{\mathscr{S}_i\}_{i=1}^M$.
For $i=M$, the covariance contribution is interpreted as zero because $\widehat{X}_M=0$ by convention; no conditional mean over $\mathscr{X}_M$ is required.

\section{Combining Runs and Redrawing}\label{app:redrawing}
In practice, it can also be beneficial to combine results from multiple independent slicing runs to improve accuracy and robustness. Each slicing run $r$ defines a valid partition of the parameter space via a sequence of slicing thresholds $\{\phi_j^{(r)}\}$, associated level sets $\{\mathscr{X}_j^{(r)}\}$, and corresponding samples $\{\Theta_{j,k}^{\mathscr{X},(r)}\}$. Multiple such runs can be combined to refine the partition and improve volume estimates by aggregating information across runs.

The combination procedure begins by forming a global list of thresholds $\{\phi_i\}$ from the union of all $\{\phi_j^{(r)}\}$, sorted in ascending order without duplication.
For each $\phi_i$, we collect all available samples from level sets corresponding to that threshold
\begin{equation}
    \{\Theta^{\mathscr{X}}_{i,k}\} =\!\!\!\! \bigcup_{(r,j): \phi^{(r)}_j = \phi_i} \!\!\!\{\Theta_{j,k}^{\mathscr{X},(r)}\}\,.
\end{equation}
This pooled sample is then used in a somewhat modified slicing algorithm (cf. Algorithm~\ref{alg:slimx})
\begin{itemize}
    \item Step 3 is modified to use the available sample $\{\Theta^{\mathscr{X}}_{i-1,k}\}$ from the previous level set $\mathscr{X}_{i-1}$.
    \item Step 4 selects a new threshold $\phi_i$ from the combined list, based on predefined criteria (e.g., the new threshold must not be too close to the previous one, such that sufficiently many points can fall within the thresholds).
\end{itemize}
If no points fall between two successive thresholds, the algorithm may skip over them. However, thresholds originating from a valid slicing run will eventually satisfy the selected slicing condition. The rest of the procedure follows the standard slicing logic, yielding updated slices, volume estimates, and filtered samples.

This approach enables reuse of already sampled points without introducing a bias. It can be used to refine the resolution of the slice partition by integrating new thresholds across runs, or redraw samples within a fixed partition to increase the number of points per slice, reducing the variance of volume estimates, and providing more samples for the intra-slice Monte Carlo estimation. In the special case where all runs share identical thresholds $\{\phi_j^{(r)}\} \equiv \{\phi_j\}$, the combination acts purely as a redrawing procedure. The samples in each slice are pooled, increasing the effective sample size $N_i = \sum_r N_i^{(r)}$, allowing higher accuracy in estimating slice volumes and expectations, as discussed in Section~\ref{sec:combining runs}.

\section{Drawing Random Points from the Posterior}\label{app:draw simple}
The goal is to draw samples from the ``expensive" target posterior PDF $p(\theta) = L(\theta) \pi(\theta)/\mathcal{Z}$ based on the rejection technique and the already stored sample $\{\Theta_{i,k}\}_{k=1}^{n_i}$ of parameter values in slice $i$, which are distributed according to the constrained prior PDF $\pi\vert_{\mathscr{S}_i}$.
To this end, we need a simple envelope PDF   
$p^{(s)}(\Theta)$ that fulfills the condition
\begin{equation}\label{eq:cond}
p(\theta) \le \frac{\kappa}{\mathcal{Z}} p^{(s)}(\theta) \;,
\end{equation}
with some suitable parameter $\kappa$ from which we can easily draw samples $\Theta^{t}$. This proposed parameter point is then accepted with probability
\begin{equation}
P_\text{acc} = \frac{L(\Theta^{t})\pi(\Theta^{t})}{\kappa p^{(s)}(\Theta^{t})} \;.
\end{equation}
For the envelope, we choose the piecewise-constant function
\begin{align*}
p^{(s)}(\theta) &= \sum_{i} a_{i} \widehat{S}_{i}  \;\pi\vert_{\mathscr{S}_i}(\theta) \springerline\approx
\sum_{i} a_{i}   \pi(\theta) \;\indicator{\mathscr{S}_i}{\theta}
\springerline\quad\text{with}\quad
\int p^{(s)}(\theta) \,\mathrm{d}V_{\theta} = \sum_{i} a_{i} \widehat{S}_{i} \overset{!}{=}1
\end{align*}
The prior volume in slice $\mathscr{S}_i$ is estimated by  $\widehat{S}_i$.
For slice $i$ the condition reads
\begin{align*}
L(\Theta) &\le \kappa a_i,
\end{align*}
according to Equation~\eqref{eq:cond}. We can ignore the normalization condition for the coefficients $a_i$ as it can be absorbed in $\kappa$.
The coefficients $a_i$ can be determined by using a small subset $I_i \subset \{\Theta_{i,k}\}_{k=1}^{n_i}$ of the predetermined parameter points in slice $i$ and we use
\begin{align*}
\kappa a_i \approx L_i^\text{max} :=\max_{\Theta\in I_i} L(\Theta) \;.
\end{align*}
The acceptance probability for a proposed parameter value, given it lies in slice $i$, then simplifies to
\begin{align*}
P_i^\text{acc} &= \frac{L(\Theta)}{L_i^\text{max}}\;.
\end{align*}
The probability that a parameter point $\Theta$ drawn from the envelope distribution falls into slice $i$ is
\begin{align}\label{eq:Pi}
P_i &= \int_{\mathscr{S}_i} p^{(s)}(\theta) \,\mathrm{d}V_\theta = \frac{L^\text{max}_i \widehat{S}_{i}}{\sum_j L^\text{max}_j \widehat{S}_{j}}\;.
\end{align}
The number of points $N_{\mathrm{post},i}$ in slice $i$ is multinomial with mean
\begin{align}\label{eq:multinomial}
\mathbb{E}[N_{\mathrm{post},i}] &= N_{\mathrm{post}} P_i\;,
\end{align}
where $N_{\mathrm{post}}$ is the total number of posterior points.
Therefore, the random numbers 
\begin{align}
D_i &:= \frac{N_{\mathrm{post},i}}{P_i}
\end{align}
have mean $N_{\mathrm{post}}$ and variance $\Var{D_i}= N_{\mathrm{post}} (1-P_i)/P_i$. The actual numbers $D_i$ can be used as a diagnostic tool for the performance of the run.

Now, drawing a parameter value $\Theta$ proceeds in two steps: first, we select a slice $i$ with probability $P_i$ and, within that slice, we draw a trial parameter point $\Theta^t$ according to the constrained prior. This parameter is accepted with probability $P_i^\text{acc}$.
Drawing a random parameter value from the constrained prior can be strongly simplified, as we already have a large sample of stored parameter points with the correct distribution. So we merely have to select uniformly from the stored subset of points belonging to slice $i$, not just those that have been used to determine $a_i$. Therefore, it may happen that the condition 
in Equation~\eqref{eq:cond} is violated. This is the case if
\begin{align*}
L(\Theta) &> L_i^\text{max}\;.
\end{align*}
A remedy would be to set $L_i^\text{max} = L(\Theta)$, but that implies that the number of points collected so far in slice $i$ is too small as compared to the number of points in the other slices (see Equations~\eqref{eq:Pi} and~\eqref{eq:multinomial}).
The ratio of the number of points in the other slices is still correct. The deficiency in slice $i$ could be cured by the {\bf greedy sampling}, i.e., by drawing selectively more points in slice $i$ until $N_{\mathrm{post},i}$ equals the mean value or until $D_i$ is equal to the minimum $D_j$ of the other slices.

But besides the number of points in the slices, also the distribution of points within slice $i$ is no longer correct, due to the change of $a_i$. Therefore, the rejection step has to be repeated in that slice with the updated envelope. Note, however, that this step does not require additional target function evaluations.

\section{Numerical Implementation Details}\label{app:numerical_implementation}

This appendix specifies how the empirical error--cost pairs used in Section~\ref{sec:experiments} are generated and how the method-specific cost mappings are defined.
The common cost variable is the number of pointwise target evaluations $Q$, but each method has its own native control parameter, and therefore a different mapping from that parameter to $Q$.

\subsection{Method setup and cost mappings}\label{app:numerical_costs}

For IS, no empirical schedule is required because $Q^\star_{\mathrm{IS}}$ is obtained analytically from the central limit theorem in Appendix~\ref{app:IS budget}.
The corresponding native control parameter is the number of importance samples $N_{\mathrm{IS}}$; each sample requires one evaluation of $L$, so $Q=N_{\mathrm{IS}}$.

For S$\ell$MC we construct the slices from the surrogate $\phi$ as in Section~\ref{sec:slicing}, with the same rule in every benchmark: the thresholds $\phi_i$ are taken as the median surrogate value of the constrained-prior samples above $\phi_{i-1}$, so every compression factor is $t_i\approx\tfrac12$ (no slice holds more than half of the samples in $\mathscr{X}_{i-1}$). Each step uses $N_i=N_\phi=10^5$ exact constrained-uniform surrogate draws in every dimension. The number of slices $M$ is an output of this construction, not a free parameter: starting from the lowest threshold, new slices are added until the running estimate of the surrogate integral stabilizes. Writing $\widehat{\mathcal{Z}}_\phi^{(i)}$ for the estimate after $i$ slices, generation stops at the first slice for which
\begin{equation}
\bigl|\log\widehat{\mathcal{Z}}_\phi^{(i)} - \log\widehat{\mathcal{Z}}_\phi^{(i-1)}\bigr| < 10^{-3},
\end{equation}
i.e.\ a relative change of the estimate below ${\sim}10^{-3}$; a final slice then collects the remaining mass. The resulting $M$ grows with dimension ($M=24,55,146$ for $d=2,8,30$), and each immutable cache contains 50,000 stored points per slice. All thresholds are fixed before any evaluations of $L$.
The target is then evaluated at $n_i$ points in slice $i$; in the reported runs, the native control parameter is the equal allocation $n_i=n$, so $Q=\sum_{i=1}^M n_i=Mn$. The floating schedule begins at $n=1$, is multiplied by 1.01 at every occurrence, and includes the exact endpoint $n=20{,}000$. Every occurrence is executed independently, including occurrences that round to the same integer $n$. This gives 997 independent occurrences at 633 distinct integer budgets.

For RW and the NS surrogate benchmark, the floating $N_{\mathrm{live}}$ schedule begins at 20, is multiplied by 1.01 at every occurrence, and includes the exact endpoint $N_{\mathrm{live}}=100{,}000$. Repeated rounded integers are retained as independent calculations, giving 857 occurrences. The Gaussian geometry permits exact constrained draws. Each run stops when the remaining-evidence diagnostic based on the geometric mean of the live likelihoods falls below a tolerance
\begin{equation}
\springeronly{\begin{aligned}&}
\frac{\exp(\langle \log L\rangle_{\mathrm{live}})\,X_i}{\widehat{\mathcal{Z}}^{(i)}}<10^{-3},
\springerline\qquad
X_i=\left(\frac{N_{\mathrm{live}}}{N_{\mathrm{live}}+1}\right)^i,
\springeronly{\end{aligned}}
\end{equation}
where $\langle \log L\rangle_{\mathrm{live}}$ denotes the arithmetic mean of the log-likelihoods of the current $N_{\mathrm{live}}$ live points and $\widehat{\mathcal{Z}}^{(i)}$ is the running evidence estimate. Dimension-specific iteration caps are safety limits and do not replace this stopping criterion. We denote the number of discarded representatives by $K$.
For RW, the target is evaluated on the $K$ discarded surrogate-NS representatives. The final $N_{\mathrm{live}}$ target values are also evaluated and included in the logical cost, but the estimator retains the discarded-representative quadrature rather than adding a live-evidence correction. Thus $Q=K+N_{\mathrm{live}}$.
The NS curve is the evidence estimate from the same isotropic surrogate benchmark, reused at every deformation value. Its logical cost is $Q=N_{\mathrm{live}}+20K$. The multiplier 20 is a benchmark convention representing constrained-replacement overhead; it is not the measured work of the exact Gaussian sampler.

Overall, the resulting target-evaluation costs are summarized as
\begin{equation}
Q =
\begin{cases}
N_{\mathrm{IS}}, & \text{IS},\\[2pt]
\sum_{i=1}^{M} n_i, & \text{S$\ell$MC},\\[2pt]
K+N_{\mathrm{live}}, & \text{RW},\\[2pt]
N_{\mathrm{live}}+20K, & \text{NS surrogate benchmark},
\end{cases}
\label{eq:cost_mappings}
\end{equation}

\subsection{Estimating the required computational budget}\label{app:numerical_nstar}
Each independent empirical occurrence contributes a realized pair $(Q_j,\delta_j)$. Candidate centres are the distinct observed costs, sorted in ascending order. Repeated occurrences at the same integer cost remain separate observations inside every window. For a multiplicative width $w=1.5$, the inclusive centred window associated with a candidate $Q_j$ is
\begin{equation}
Q\in\left[\frac{Q_j}{w},\,wQ_j\right].
\end{equation}
The candidate qualifies when
\begin{equation}
\frac{\#\{k:\delta_k\le\epsilon,\;Q_k\in[Q_j/w,wQ_j]\}}
{\#\{k:Q_k\in[Q_j/w,wQ_j]\}}
\ge p.
\end{equation}
with $p=0.9$. The first qualifying centre is reported as $Q^\star$. If $Q_j/w$ lies below the minimum simulated cost, the available observations are used; no unobserved failures or successes are inserted. Complete upper coverage is mandatory, so a candidate is considered only when the original scan reaches at least $wQ_j$. Bootstrap resamples retain these original coverage limits.

We draw 2,000 bootstrap samples of the complete empirical point set and repeat the same detection. For comparisons based on shared random streams, the same resampling indices are used for the paired methods. We report the median and 16th/84th percentiles of successful detections. A result is accepted for plotting when more than ten bootstrap resamples succeed and the successful fraction is at least 0.8. Failed replicates are retained as failures when calculating this support. Consequently, an omitted point means that the simulated range did not support a sufficiently stable convergence estimate under this criterion.

The main $d=2$ scan uses anisotropies $a=1.2,2.4,\ldots,13.2$ and 11 equally spaced shifts from $b=0.2$ to $17.755556$. The $d=8$ scan adds exact agreement, $a=1.0$ and $b=0.0$, to 13 equally spaced anisotropies from $a=1.2$ to $2.6181818$ and 13 equally spaced shifts from $b=0.2$ to $4.3454545$. The $d=30$ scan likewise adds $a=1.0$ and $b=0.0$ to 12 equally spaced anisotropies from $a=1.2$ to $2.5$ and 12 equally spaced shifts from $b=0.2$ to $4.0$. The control-variate comparison uses $a=1.0,1.1$ and $b=0.0,0.1$ in both $d=2$ and $d=8$.

\section{A Control Variate for Slice Monte Carlo}\label{app:slmc_variants}

The surrogate values already stored at the S$\ell$MC points can also be used as a control variate. For slice $\mathscr S_i$, define
\begin{equation}
\widehat{\mathcal Z}_{L,i}
=
\frac{1}{n_i}\sum_{k=1}^{n_i}L(\Theta^{\mathscr S}_{i,k}),
\qquad
\widehat{\mathcal Z}_{\phi,i}
=
\frac{1}{n_i}\sum_{k=1}^{n_i}\phi(\Theta^{\mathscr S}_{i,k}).
\end{equation}
Let $\mathcal Z^{\mathrm{ref}}_{\phi,i}$ denote an accurately determined reference mean of the surrogate in slice $i$. The control-variate estimator with one global coefficient $\alpha$ shared by all slices is
\begin{equation}\label{eq:paired_global_alpha_estimator}
\widehat{\mathcal Z}^{(\alpha)}
=
\sum_{i=1}^{M}\widehat S_i
\left[
\widehat{\mathcal Z}_{L,i}
-\alpha\left(
\widehat{\mathcal Z}_{\phi,i}
-\mathcal Z^{\mathrm{ref}}_{\phi,i}
\right)
\right].
\end{equation}
For $\alpha=0$ this is ordinary S$\ell$MC. With the stratification, volume estimates, reference means and sample counts held fixed, the common coefficient minimizing the variance of Equation~\eqref{eq:paired_global_alpha_estimator} under independent sampling is
\begin{equation}\label{eq:paired_global_alpha_optimal}
\alpha^\star
=
\frac{
\displaystyle\sum_{i=1}^{M}\frac{\widehat S_i^2}{n_i}
\Covt{L(\Theta),\phi(\Theta)}{\pi\vert_{\mathscr S_i}}
}{
\displaystyle\sum_{i=1}^{M}\frac{\widehat S_i^2}{n_i}
\Vart{\phi(\Theta)}{\pi\vert_{\mathscr S_i}}
}.
\end{equation}

Separate coefficients could be fitted in each slice, but reliable slice-wise fits can require more target evaluations.

In the numerical comparison, the moments in Equation~\eqref{eq:paired_global_alpha_optimal} are replaced by their same-sample empirical counterparts. The same observations are therefore used both to fit $\widehat\alpha$ and to evaluate Equation~\eqref{eq:paired_global_alpha_estimator}. This plug-in construction can have finite-sample bias; an independent pilot estimate or cross-fitting would be required to remove the same-sample fitting effect in a production calculation.

The surrogate reference mean $\mathcal Z^{\mathrm{ref}}_{\phi,i}$ is estimated from the full immutable pool of 50,000 stored surrogate points in each slice. When $n_i=1$, or when the selected surrogate values have exactly zero variance, the implementation uses the fallback $\widehat\alpha=1$. Signed integral estimates are retained.

The comparison uses the completed $d=2$ and $d=8$ main-study slice caches. At each shared mismatch and budget, both estimators use the same sample points and target evaluations. Both estimators have target-evaluation cost $Q=Mn$, where $M$ is the number of slices and $n$ the number of target evaluations per slice; surrogate evaluations are not included in $Q$.

The points-per-slice schedule contains 997 independent occurrences from $n=1$ through $n=20{,}000$, generated with floating factor 1.01 and retaining repeated rounded integer budgets. The study adds the near-agreement coordinates $a=1.0,1.1$ and $b=0.0,0.1$ to the main target grids. The same centred-window and bootstrap procedure as in Appendix~\ref{app:numerical_nstar} is used. Both dimensions use $\epsilon=0.1$, $p=0.9$, $w=1.5$, 2,000 paired bootstrap resamples, and a minimum support of 80\%. The bootstrap uses identical resampling indices for ordinary and control-variate S$\ell$MC.

\begin{figure*}[tb]
  \centering
  \includegraphics[width=\textwidth]{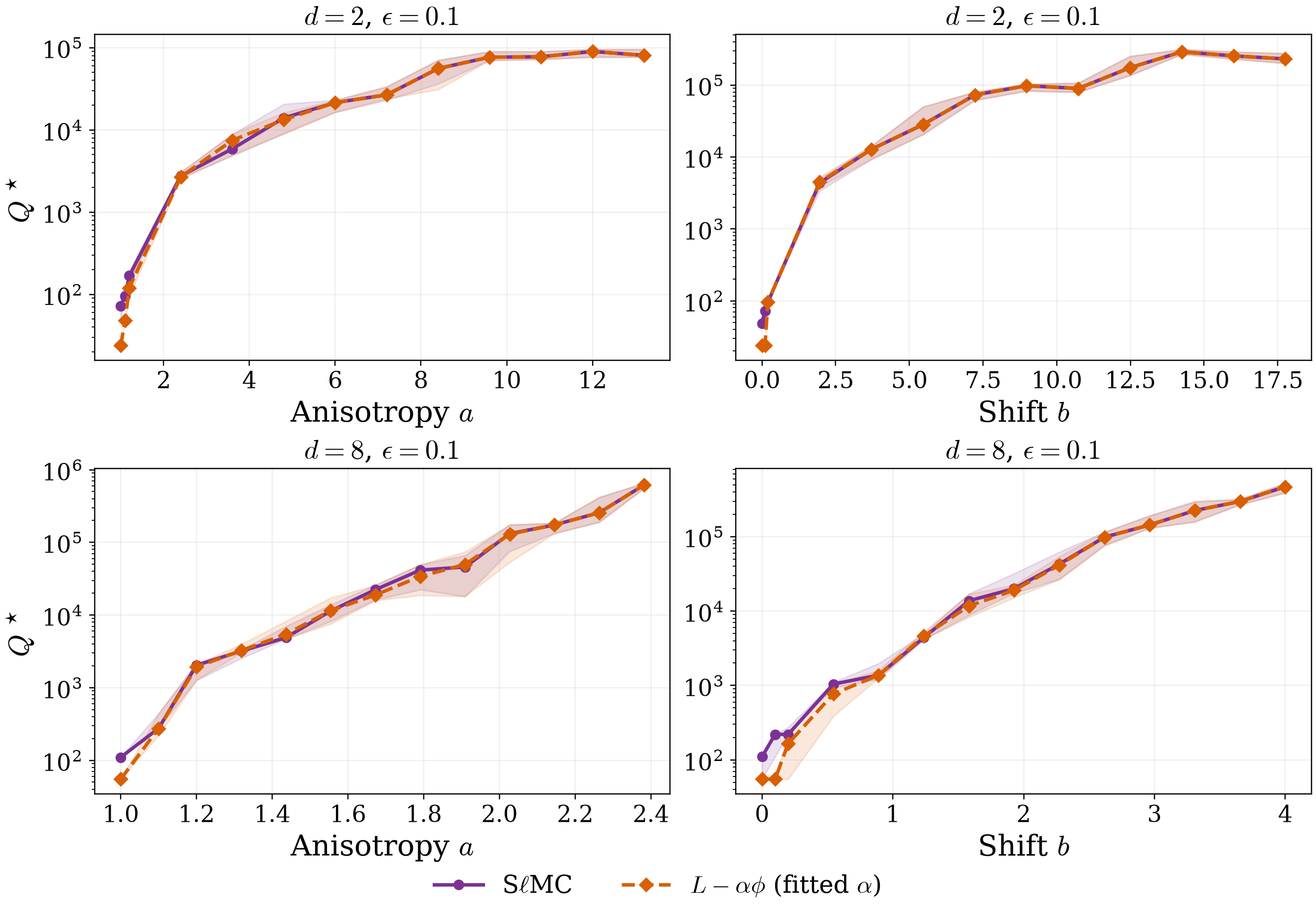}
  \caption{Required target-evaluation cost $Q^\star$ for ordinary S$\ell$MC and control-variate S$\ell$MC with an empirically fitted global coefficient $\widehat\alpha$. The top row shows $d=2$ and the bottom row $d=8$; the left column varies anisotropy $a$ and the right column varies shift $b$. All panels use $\epsilon=0.1$, $p=0.9$, and centred windows $[Q^\star/1.5,1.5Q^\star]$. Available lower-bound observations are used, complete upper coverage is required, and repeated integer-budget occurrences remain independent observations. Lines show bootstrap medians and shaded regions show 16th/84th percentiles from 2,000 paired bootstrap resamples. Results with less than 80\% bootstrap support are omitted.}
  \label{fig:slmc_variants}
\end{figure*}

At exact surrogate--target agreement, the control variate removes the within-slice target variation and reaches the minimum equal-allocation cost of one target evaluation per slice: $Q^\star=24$ in $d=2$, and the corresponding near-agreement cases can reach $Q^\star=55$ in $d=8$. A reduction is also visible for the nearest nonzero mismatches.

The benefit decays rapidly as the surrogate--target mismatch grows. Away from agreement, the fitted control-variate curve largely overlaps ordinary S$\ell$MC and does not uniformly reduce target cost. This reflects the stronger information required by the control variate: standard S$\ell$MC needs the surrogate-induced ordering to remain informative, whereas the control variate additionally requires strong pointwise correlation between $L$ and $\phi$ within each slice. Because $\widehat\alpha$ is fitted on the same observations used in the final estimate, occasional increases in empirical cost are compatible with finite-sample fitting noise and possible plug-in bias. We therefore regard the comparison as a diagnostic demonstration of the control-variate opportunity, not as evidence that the fitted estimator uniformly replaces ordinary S$\ell$MC.

\section{Analytical Importance Sampling Budget for Gaussian Benchmarks}\label{app:IS budget}
Using the IS notation from Section~\ref{sec:IS}, the analytical IS budget follows directly from the central limit theorem,
\begin{equation}
\sqrt{N}\,
\frac{\widehat{\mathcal{Z}}_N^\text{IS}-\mathcal{Z}}
{\sigma_{\mathrm{IS}}}
\xrightarrow{d}
\mathcal{N}(0,1)\,,
\end{equation}
with asymptotic standard deviation $\sigma_{\mathrm{IS}} = \mathcal{Z}_\phi\sqrt{\mathrm{\mathbb{V}ar}_{p_\phi}[L/\phi]}$. For a relative error threshold $\epsilon$ and success probability $p$, this gives
\begin{equation}
Q^\star_{\mathrm{IS}}
\approx
\left(
\frac{z_{(1+p)/2}}{\epsilon}
\frac{\sigma_{\mathrm{IS}}}{\mathcal{Z}}
\right)^{\!2}\,,
\end{equation}
where $z_{(1+p)/2}$ is the $(1+p)/2$-quantile of the standard normal distribution.

For a diagonal Gaussian surrogate and target, the normalized variance factor in this expression has a fully closed form in terms of relative shifts and width ratios.
Let
\begin{equation}
\Delta_i := \frac{\mu_{L,i}-\mu_{\phi,i}}{\sigma_{L,i}},
\qquad
\tau_i := \frac{\sigma_{\phi,i}}{\sigma_{L,i}}.
\end{equation}
A direct Gaussian integral gives
\begin{equation}
\springeronly{\begin{aligned}&}
\frac{\mathcal{Z}_\phi^2\,\mathrm{\mathbb{V}ar}_{p_\phi}[L/\phi]}{\mathcal{Z}^2}
\springerline=
\prod_{i=1}^{d}
\frac{\tau_i^2}{\sqrt{2\tau_i^2-1}}
\exp\!\left(\frac{\Delta_i^2}{2\tau_i^2-1}\right)
-1.
\label{eq:IS_normalized_variance_general}
\springeronly{\end{aligned}}
\end{equation}
This is finite if and only if $\tau_i^2>1/2$ in every dimension, for each $i=1,\ldots,d$.

\emph{Shift family} ($\tau_i=1$, $\Delta_i\neq0$):
Setting $\tau_i=1$ gives
\begin{equation}
\frac{\mathcal{Z}_\phi^2\,\mathrm{\mathbb{V}ar}_{p_\phi}[L/\phi]}{\mathcal{Z}^2}
=
\exp\!\left(\sum_{i=1}^d \Delta_i^2\right)-1.
\end{equation}
This grows exponentially in the squared shift, explaining the catastrophic degradation of IS under mean mismatch.

\emph{Anisotropy family} ($\Delta_i=0$):
Setting $\Delta_i=0$ removes the exponential factor, giving
\begin{equation}
\frac{\mathcal{Z}_\phi^2\,\mathrm{\mathbb{V}ar}_{p_\phi}[L/\phi]}{\mathcal{Z}^2}
=
\prod_{i=1}^d
\frac{\tau_i^2}{\sqrt{2\tau_i^2-1}}
-1,
\end{equation}
which is finite only when $\tau_i^2>1/2$ for each $i=1,\ldots,d$.
In the main-text anisotropy parameterization, the binding constraint comes from the stretched half, for which $\tau_i=1/a$; finite variance therefore requires $a<\sqrt{2}$.
The compressed half ($\tau_i=a$) satisfies the condition for all $a\ge 1$.
At $a=\sqrt{2}$, the normalized variance diverges and $Q^\star_{\mathrm{IS}}\to\infty$.
    
\end{appendix}

\begin{acknowledgments}
The authors gratefully acknowledge support from Graz University of Technology.
\end{acknowledgments}

\section*{Data availability}
The data supporting the findings of this study are available from the corresponding author upon reasonable request.

\bibliography{main}
\end{document}